\documentclass[aps,twocolumn]{revtex4}
\usepackage[colorlinks, linkcolor=red,anchorcolor=green,citecolor=blue]{hyperref}
\usepackage{graphicx}
\usepackage{bm}
\usepackage{amsmath}
\usepackage{color}
\usepackage{booktabs}
\usepackage{multirow}
\usepackage{diagbox}

\begin{document}
\title{Fully-heavy tetraquarks in strongly interacting medium }
\author{Jiaxing Zhao$^a$}
\author{Shuzhe Shi$^b$}
\email{shuzhe.shi@mcgill.ca}
\author{Pengfei Zhuang$^a$}
\affiliation{$^a$Physics Department, Tsinghua University, Beijing 100084, China\\
      	 $^b$Department of Physics, McGill University, Montreal, QC H3A 2T8, Canada}
\date{\today}
\begin{abstract}
We study the properties of fully-heavy tetraquarks at finite temperature and their production in high-energy nuclear collisions. We obtain the masses and wave functions of the exotic hadron states $cc\bar c\bar c$ and $bb\bar b\bar b$ by solving the four-body Schr\"odinger equation in vacuum and strongly interacting matter. In vacuum, the tetraquarks are above the corresponding meson-meson mass threshold, and the newly observed exotic state $X(6900)$ might be a $cc\bar c\bar c$ state with quantum number $J^{PC}=0^{++}$ or $1^{+-}$. In hot medium, the temperature dependence of the tetraquark masses and the dissociation temperatures are calculated. Taking the wave function at finite temperature, we construct the Wigner function for the tetraquark states and calculate, with coalescence mechanism, the production yield and transverse momentum distribution of $cc\bar c\bar c$ in heavy-ion collisions at LHC energy. In comparison with nucleon-nucleon collisions, the yield per binary collision is significantly enhanced.  
\end{abstract}

\maketitle

\section{introduction}
\label{section1}
The quantum chromodynamics (QCD), which is widely accepted as the theory of strong interaction, allows the existence of exotic hadrons, such as glueballs containing only gluons~\cite{Mathieu:2008me,Ochs:2013gi}, hybrids with quarks and gluons~\cite{Meyer:2015eta,Chanowitz:1982qj}, multi-quark states like tetraquarks and pentaquarks~\cite{Esposito:2016noz,Karliner:2017qhf} and hadronic molecules~\cite{DeRujula:1976zlg,Guo:2017jvc}. There are a lot of candidates for exotic hadrons in the light-quark sector, such as $a_0(980)$, $f_0(1370)$ and $\Lambda(1405)$. In 2003, the Belle Collaboration discovered a new hadron state, named $X(3872)$~\cite{Choi:2003ue}. It cannot be explained as a normal meson or a baryon, since its decay properties indicate that it contains a pair of charm quarks. This is the first discovery of exotic hadrons with heavy quarks. After that, many more hadrons are found in processes with final states containing a heavy quark-antiquark pair, and such hadrons are refereed to as XYZ states. So far, there have been more than thirty XYZ states discovered in experiments, see recent reviews \cite{Guo:2017jvc,Richard:2016eis,Hosaka:2016pey,Ali:2017jda,Liu:2019zoy}. Among the studies, there are many theoretical works focusing on fully-heavy tetraquarks $QQ\bar Q\bar Q$ ($Q=c,b$)~\cite{Badalian:1985es,Ader:1981db,Berezhnoy:2011xn,Zouzou:1986qh,Brink:1998as,Karliner:2016zzc,Debastiani:2017msn,Wang:2019rdo,Liu:2019zuc,Chen:2020lgj,Yang:2020rih,Lu:2020cns}. The advantage of studying fully-heavy tetraquarks is the nonrelativistic treatments which largely simplify the calculations. Such treatments include lattice QCD~\cite{Bicudo:2015vta,Bicudo:2017usw}, QCD sum rules~\cite{Chen:2016jxd,Wang:2017jtz} and potential models~\cite{Brink:1998as,Debastiani:2017msn,Wang:2019rdo,Chen:2020lgj,Yang:2020rih,Lu:2020cns}. Recently, a narrow structure around $6.9$~GeV, named $X(6900)$, is observed by the LHCb Collaboration at $\sqrt{s}=7, 8, 13$~TeV~\cite{Aaij:2020fnh}. This is the first candidate of fully-heavy tetraquarks observed in experiment. 

The main difficulty of observing fully-heavy tetraquarks in elementary collisions, such as electron-positron and nucleon-nucleon collisions, is the small production cross section of heavy quarks. The formation of a fully-heavy tetraquark requires at least two pairs of heavy quarks with small relative momenta, which is very rare in an elementary event. This difficulty can be overcame in high-energy nuclear collisions. Since the binding energy among the nucleons of a nucleus can be safely neglected at high energies, a nucleus-nucleus collision contains a number of binary nucleon-nucleon collisions. Therefore, the number of heavy quarks and in turn the number of fully-heavy tetraquarks will be significantly enhanced in high-energy nuclear collisions. From the experimental data~\cite{Adamczyk:2014uip,Abelev:2012vra}, the charm quark number can reach $10$ at the Relativistic Heavy-Ion Collider (RHIC) and even $100$ at the Large Hadron Collider (LHC). After the creation in the initial stage of the collisions, the heavy quarks will pass through the new state of matter of light quarks and gluons which is called the quark-gluon plasma (QGP). Due to the strong interaction with the matter, heavy quarks are widely considered as a sensitive probe of the QGP~\cite{Bedjidian:2004gd,Zhao:2020jqu}. The energy loss of heavy quarks during the evolution in the hot medium makes them be partially or even fully thermalized with the matter before the hadronization. Finally, on the hadronization hypersurface of heavy quarks, tetraquark states are formed via coalescence mechanism~\cite{Fries:2003kq,Molnar:2003ff,Lin:2002rw,Hwa:2002tu,Oh:2009zj}. The key factor in all coalescence models for light and heavy hadrons is the coalescence probability for quarks to form a hadron state. Considering the big problem of confinement, the coalescence probability or the Wigner function is normally taken as a Gaussian distribution with adjustable parameters~\cite{Fries:2003kq,Molnar:2003ff,Lin:2002rw,Hwa:2002tu,Oh:2009zj}.          

Taking into account the fact that charm and bottom quarks are very heavy and their moving velocity is small, there exists a hierarchy of scales in the study of heavy quarks: $m\gg mv \gg mv^2$~\cite{Caswell:1985ui,Brambilla:1999xf}. Integrating out the degrees of freedom with momenta larger than $m$ and $mv$ successively in the QCD Lagrangian, one can derive its nonrelativistic versions NRQCD and pNRQCD~\cite{Brambilla:1999xf}. Furthermore, if neglecting the interaction between color-singlet and color-octet states, the pNRQCD becomes a potential model~\cite{Brambilla:1999xf}. In this case, one can employ the Schr\"odinger equation to study the properties of hadrons consist of only heavy quarks. The potential model has been successfully applied to open and closed heavy flavors in vacuum and at finite temperature~\cite{Satz:2005hx,Zhao:2017gpq,Shi:2019tji}. It is pointed out that, in comparison with nucleon-nucleon collisions in vacuum the $\Xi_{cc}$ and $\Omega_{ccc}$ yields per binary nucleon-nucleon collision in heavy-ion collisions at RHIC and LHC will be largely enhanced~\cite{Zhao:2016ccp,He:2014tga}. In this work, we employ the four-body Schr\"odinger equation to study the properties of fully-heavy tetraquark states $cc\bar c\bar c$ and $bb\bar b\bar b$ at finite temperature and their production in high-energy nuclear collisions. Different from light hadrons where the coalescence probability is assumed to be a Gaussian distribution, the probability for fully-heavy tetraquarks is derived from the wave function of the system controlled by the Schr\"odinger equation. This is essential for predicting the properties of unconfirmed particles.

The structure of the paper is as follows. In Sec. \ref{section2} we present the theoretical framework of solving the four-body Schr\"odinger equation. The tetraquark properties, including mass and size, in vacuum and hot medium are investigated in Secs. \ref{section3} and \ref{section4}. In Sec. \ref{section5} the total yield and transverse momentum distribution of the fully-charmed tetraquark state $cc\bar c\bar c$ in heavy-ion collisions are calculated and compared with its production in nucleon-nucleon collisions. After summarizing in Sec. \ref{section6}, we provide supplementary informations about the hyperspherical harmonic functions in Appendix \ref{appendix1}, and the method to compute the potentials is described in Appendix \ref{appendix2}.

\section{Theoretic framework}\label{sec.framework}
\label{section2}
For a system of four quarks with the same mass $m$, the wave function $\Psi({\bf r}_1,{\bf r}_2,{\bf r}_3,{\bf r}_4)$ and the energy $E$ are controlled by the Schr\"odinger equation
\begin{equation}
\left( \sum_{i=1}^4 {\widehat {\bf q}^2_i\over 2m}+ \sum_{i<j}V_{ij}(|{\bf r}_{ij}|)\right)\Psi = E\Psi,
\end{equation}
where we have assumed that the interaction potential $V=\sum_{i<j}V_{ij}$ is the summation of the two-body interactions, and the direct three- and four-body potentials are neglected. Taking into account one-gluon-exchange interaction, the two-body potential can be effectively expressed as~\cite{Wong:2001td,Kawanai:2011jt},
\begin{equation}
V_{ij}(|{\bf r}_{ij}|) = -{1\over 4}\lambda_i^a\cdot\lambda_j^a\left(V^c_{ij}(|{\bf r}_{ij}|)+V^{ss}_{ij}(|{\bf r}_{ij}|){\bf s}_i\cdot{\bf s}_j\right),  
\label{v2}
\end{equation}
where $\lambda_i^a\ (a=1,...,8)$ are the SU(3) Gell-Mann matrices, the factor $1/4$ is from the normalization, $V^c_{ij}$ is the spin independent interaction, $V^{ss}_{ij}$ is the strength of the spin-spin interaction, and $|{\bf r}_{ij}|=|{\bf r}_i-{\bf r}_j|$ is the distance between the two quarks labeled by $i$ and $j$. In order to solve the four-body Schr\"odinger equation, we first introduce the Jacobi coordinates,
\begin{eqnarray}
{\bf X} &=& {1\over 4}({\bf r}_1+{\bf r}_2+{\bf r}_3+{\bf r}_4) \nonumber\\
{\bf x}_1&=&\sqrt{3m\over 4\mu}\left({\bf r}_4-{{\bf r}_1+{\bf r}_2+{\bf r}_3\over 3}\right)\nonumber\\
{\bf x}_2&=&\sqrt{2m\over 3\mu}\left({\bf r}_3-{{\bf r}_1+{\bf r_2}\over 2}\right)\nonumber\\
{\bf x}_3&=&\sqrt{m\over 2\mu}\left({\bf r}_2-{\bf r}_1\right),
\label{jacobi}
\end{eqnarray}
where $\mu$ is a parameter with dimension of mass and its value does not affect the final result~\cite{Krivec:1998}. We take $\mu=M=4m$ in numerical calculations. With such coordinates, the kinetic energy becomes
\begin{equation}
\sum_{i=1}^4 {\widehat {\bf q}^2_i\over 2m} = \frac{{\bf P}^2}{2M}+\frac{{\bf p}_1^2}{2\mu} + \frac{{\bf p}_2^2}{2\mu} + \frac{{\bf p}_3^2}{2\mu}.
\end{equation}

Since the potential depends only on the relative coordinates ${\bf x}_i$, one can factorize the four-body motion into a center-of-mass motion and a relative motion,  $\Psi({\bf r}_1,{\bf r}_2,{\bf r}_3,{\bf r}_4)=\Theta({\bf X})\Phi({\bf x}_1,{\bf x}_2,{\bf x}_3)$. The bound state properties only relate to the relative motion of the system, and we just need to deal with the nine dimensional wave equation. We then express the relative coordinates ${\bf x}_1$, ${\bf x}_2$ and ${\bf x}_3$ in the hyperspherical frame: hyperradius $\rho=\sqrt{x_1^2+x_2^2+x_3^2}$ and hyperangles $\Omega=\{\alpha_2,\alpha_3, \theta_1,\phi_1, \theta_2,\phi_2,\theta_3,\phi_3\}$, where the angles $\alpha_2\equiv\arcsin(x_2 /\sqrt{x_1^2+x_2^2})$ and $\alpha_3 \equiv \arcsin(x_3/ \rho)$ are defined within the range $[0,\pi/2]$, and $\{x_i, \theta_i, \phi_i\}$ are the spherical coordinates corresponding to ${\bf x}_i$. With the hyperspherical coordinates, the Schr\"odinger equation governing the relative wave function $\Phi(\rho,\Omega)$ can be written as 
\begin{eqnarray}
\left[ {1\over 2\mu}\left( -{d^2 \over d\rho^2}-{8\over \rho}{d \over d\rho}  + {\widehat {\bf K}_3^2\over \rho^2}\right) + V(\rho, \Omega) \right ]\Phi = E_r \Phi,
\label{relative}
\end{eqnarray}
where $\widehat {\bf K}_3$ is the hyperangular momentum operator, and $E_r$ the relative energy (binding energy).

As shown in \eqref{v2}, the potential $V(\rho, \Omega)$ depends on the color and spin degrees of freedom. We start with constructing the color and spin sector of the wave-function, based on the symmetry properties. We follow the analysis in Ref.~\cite{Park:2013fda}. The Pauli exclusion principle requires the wave-function to be anti-symmetric when exchanging two identical fermions, i.e. two quarks or two antiquarks. From the decomposition in color space, 
\begin{eqnarray}
(3_c\otimes 3_c)\otimes (\bar 3_c \otimes \bar 3_c) &=& \bar 3_c \otimes 3_c \oplus  6_c \otimes \bar 6_c\nonumber\\
&\oplus&  \bar 3_c \otimes \bar 6_c \oplus   6_c \otimes 3_c,
\end{eqnarray}
there are two color-singlet states obtained from the first and second terms on the right-hand side. We label them as  
\begin{eqnarray}
|\phi_1\rangle &=& |(QQ)_{\bar 3_c}(\bar Q \bar Q)_{3_c} \rangle,\nonumber\\
|\phi_2\rangle &=& |(QQ)_{6_c}(\bar Q \bar Q)_{\bar 6_c} \rangle.
\end{eqnarray}
For the exchange between the two quarks or two antiquarks, $|\phi_1\rangle$ is antisymmetric and $|\phi_2\rangle$ symmetric. 

The decomposition in spin-space is
\begin{equation}
2\otimes 2\otimes 2 \otimes 2=1 \otimes 1 \oplus  1 \otimes 3 \oplus  3 \otimes 1  \oplus   3 \otimes 3.
\end{equation}
There are two $s=0$ states,
\begin{eqnarray}
|\chi_1\rangle &=& |(QQ)_0(\bar Q \bar Q)_0 \rangle_0,\nonumber\\
|\chi_2\rangle &=& |(QQ)_1(\bar Q \bar Q)_1 \rangle_0,
\end{eqnarray}
three $s=1$ states,
\begin{eqnarray}
|\chi_3\rangle &=& |(QQ)_0(\bar Q \bar Q)_1 \rangle_1,\nonumber\\
|\chi_4\rangle &=& |(QQ)_1(\bar Q \bar Q)_0 \rangle_1,\nonumber\\
|\chi_5\rangle &=& |(QQ)_1(\bar Q \bar Q)_1 \rangle_1,
\end{eqnarray}
and one $s=2$ state
\begin{equation}
|\chi_6\rangle=|(QQ)_1(\bar Q \bar Q)_1 \rangle_2,
\end{equation}
where the subscripts denote the spin of the subsystems $QQ$ and $\bar Q\bar Q$ and the whole system $QQ\bar Q\bar Q$.
 
As we focus on the tetraquark states consist of identical quarks and antiquarks, the flavor wave function is symmetric by definition. As a first step, we consider the states with vanishing orbital angular momentum, the space wave function is then symmetric. The Pauli exclusion principle only allows the following combination of color and spin wave functions: $|\phi_1\chi_2\rangle$ and $|\phi_2\chi_1\rangle$ for the states with $J^{PC}=0^{++}$, $|\phi_1\chi_5\rangle$ for $J^{PC}=1^{+-}$, and $|\phi_1\chi_6\rangle$ for $J^{PC}=2^{++}$. While these states are orthogonal to each other, the matrix element $\langle \phi_2\chi_1| (\lambda_i^a\cdot \lambda_j^a)({\bf s}_i\cdot {\bf s}_j) |\phi_2\chi_1\rangle = -\sqrt{3/2}$ should be particularly noted in the calculation of the potential.
 
For the tetraquark states with $J^{PC}=0^{++}$, the color-spin wave function is a mixture of $|\phi_1\chi_2\rangle$ and $|\phi_2\chi_1\rangle$, and the potential contains diagonal and off-diagonal elements in color-spin space, 
\begin{eqnarray}
V_1 &=& \langle\phi_1\chi_2| \sum_{i<j} V_{ij}|\phi_1\chi_2 \rangle\nonumber\\
    &=& {2\over 3}\left(V^c_{12}+V^c_{34}\right)+{1\over 3}\left(V^c_{13}+V^c_{14}+V^c_{23}+V^c_{24}\right)\nonumber\\ 
    &+&	{1\over 6}\left(V^{ss}_{12}+V^{ss}_{34}\right)-{1\over 6}\left(V^{ss}_{13}+V^{ss}_{14}+V^{ss}_{23}+V^{ss}_{24}\right),\nonumber\\
V_2 &=& \langle\phi_2\chi_1| \sum_{i<j} V_{ij} |\phi_2\chi_1 \rangle\nonumber\\
    &=& -{1\over 3}\left(V^c_{12}+V^c_{34}\right)+{5\over 6}\left(V^c_{13}+V^c_{14}+V^c_{23}+V^c_{24}\right)\nonumber\\ 
    &+& {1\over 4}\left(V^{ss}_{12}+V^{ss}_{34}\right),\nonumber\\
V_m &=& \langle\phi_1\chi_2| \sum_{i<j} V_{ij} |\phi_2\chi_1 \rangle\nonumber\\ 
    &=&	\langle\phi_2\chi_1| \sum_{i<j} V_{ij} |\phi_1\chi_2 \rangle\nonumber\\
    &=& -{\sqrt{6}\over 8}\left(V^{ss}_{13}+V^{ss}_{14}+V^{ss}_{23}+V^{ss}_{24}\right),
\end{eqnarray}
where for the potentials $V_{ij}^c$ and $V^{ss}_{ij}$ we have explicitly labeled the two quarks with indices $i,j=1,2$ and the two antiquarks with $i,j=3,4$.

For the states with $J^{PC}=1^{+-}$ or $J^{PC}=2^{++}$, the color-spin wave functions are the eigenstates of both $(\lambda_i^a\cdot \lambda_j^a)$ and $(\lambda_i^a\cdot \lambda_j^a)({\bf s}_i\cdot {\bf s}_j)$, and the corresponding potentials are 
\begin{eqnarray}
V &=& \langle\phi_1\chi_5| \sum_{i<j} V_{ij} |\phi_1\chi_5 \rangle\\
  &=& {2\over 3}\left(V^c_{12}+V^c_{34}\right)+{1\over 3}\left(V^c_{13}+V^c_{14}+V^c_{23}+V^c_{24}\right)\nonumber\\
  &+& {1\over 6}\left(V^{ss}_{12}+V^{ss}_{34}\right)-{1\over 12}\left(V^{ss}_{13}+V^{ss}_{14}+V^{ss}_{23}+V^{ss}_{24}\right),\nonumber
\end{eqnarray}
and
\begin{eqnarray}
V &=& \langle\phi_1\chi_6| \sum_{i<j} V_{ij} |\phi_1\chi_6 \rangle\\
  &=& {2\over 3}\left(V^c_{12}+V^c_{34}\right)+{1\over 3}\left(V^c_{13}+V^c_{14}+V^c_{23}+V^c_{24}\right)\nonumber\\
  &+& {1\over 6}\left(V^{ss}_{12}+V^{ss}_{34}\right)+{1\over 12}\left(V^{ss}_{13}+V^{ss}_{14}+V^{ss}_{23}+V^{ss}_{24}\right).\nonumber
\end{eqnarray}

The potential $V(\rho, \Omega)$ depends not only on the hyperradius but also the eight hyperangles. In this case, the Schr\"odinger equation~\eqref{relative} cannot be further factorized into a radial part and an angular part. Instead, one expands the wave function in terms of the hyperspherical harmonic functions ${\mathcal Y}_\kappa(\Omega)$ which are the eigenstates of the hyperangular momentum operator $\widehat {\bf K}_3^2$,
\begin{equation}
\widehat {\bf K}_3^2{\mathcal Y}_\kappa(\Omega)=K(K+7){\mathcal Y}_\kappa(\Omega),
\end{equation}
where $\kappa$ stands for all the quantum numbers related to the angels, and $K$ is the quantum number describing the magnitude of the angular momentum. Some properties of the hyperspherical harmonic functions ${\mathcal Y}_\kappa(\Omega)$ which will be used in the following calculation are presented in Appendix~\ref{appendix1}, and the details can be found in Refs.~\cite{Krivec:1998,Barnea:1999be,Barnea:2006sd}. 

With the above preparations, we now write down the relative wave functions
\begin{eqnarray}
\Phi(\rho, \Omega) &=& \sum_\kappa\big[R^{(1)}_\kappa(\rho){\mathcal Y}_\kappa(\Omega)|\phi_1\chi_{2}\rangle\nonumber\\
                   &+& R^{(2)}_\kappa(\rho){\mathcal Y}_\kappa(\Omega)|\phi_2\chi_{1}\rangle\big]
\end{eqnarray}
for the $0^{++}$ states, 
\begin{equation}
\Phi(\rho, \Omega) = \sum_\kappa R_\kappa(\rho){\mathcal Y}_\kappa(\Omega)|\phi_1\chi_{5}\rangle
\end{equation}
for the $1^{+-}$ states and
\begin{equation}
\Phi(\rho, \Omega) = \sum_\kappa R_\kappa(\rho){\mathcal Y}_\kappa(\Omega)|\phi_1\chi_{6}\rangle
\end{equation}
for the $2^{++}$ states, where $R_\kappa(\rho)$ is the radial wave function corresponding to the hyperspherical harmonic function ${\mathcal Y}_\kappa(\Omega)$. Substituting the above expansions into the relative equation \eqref{relative}, one obtains the coupled radial equations,
\begin{eqnarray} 
&& -{1\over 2\mu}\left({d^2 \over d\rho^2}+{8\over \rho}{d \over d\rho} - {K(K+7)\over \rho^2}\right) R_\kappa\nonumber\\
&& +\sum_{\kappa'}V^{\kappa \kappa'}R_{\kappa'}= E_r R_{\kappa}
\label{radial1}
\end{eqnarray}
for $1^{+-}$ and $2^{++}$ states, and
\begin{eqnarray}
&& -{1\over 2\mu}\left({d^2 \over d\rho^2}+{8\over \rho}{d \over d\rho} - {K(K+7)\over \rho^2}\right)R^{(1)}_{\kappa}\nonumber\\ 
&& +\sum_{\kappa'}V_1^{\kappa \kappa'}R^{(1)}_{\kappa'} +\sum_{\kappa'}V_m^{\kappa \kappa'}R^{(2)}_{\kappa'}= E_r R^{(1)}_{\kappa},\nonumber \\
&& -{1\over 2\mu}\left({d^2 \over d\rho^2}+{8\over \rho}{d \over d\rho} - {K(K+7)\over \rho^2}\right)R^{(2)}_{\kappa}\nonumber\\ 
&& +\sum_{\kappa'}V_2^{\kappa \kappa'}R^{(2)}_{\kappa'}	+\sum_{\kappa'}V_m^{\kappa \kappa'}R^{(1)}_{\kappa'}= E_r R^{(2)}_{\kappa}
\label{radial2}
\end{eqnarray}
for $0^{++}$ states, where $V^{\kappa \kappa'}$ is the potential matrix element in angular momentum space
\begin{equation}
\label{vmatrix}
V^{\kappa \kappa'}=\int V(\rho, \Omega) {\mathcal Y}_\kappa^*(\Omega) {\mathcal Y}_{\kappa'}(\Omega) d\Omega \,,
\end{equation}
with the volume element
\begin{eqnarray}
d\Omega &=& \cos^5\alpha_3 \sin^2 \alpha_3  \cos^2 \alpha_2 \sin^2 \alpha_2 \sin \theta_1 \sin \theta_2 \sin \theta_3\nonumber\\
        &\times& d\alpha_3 d\alpha_2 d\theta_1 d\theta_2 d\theta_3 d\phi_1 d\phi_2 d\phi_3.
\end{eqnarray}
It is worth noting that computing the potential matrix is nontrivial. In the most general form, \eqref{vmatrix} is an eight dimensional integral, which is computationally expensive. However, taking the assumption that the total interaction potential is the summation of two-body interaction $V_{ij}(|{\bf r}_{ij}|)$, one can reduce \eqref{vmatrix} into a one dimensional integral by performing particle permutation. We show the details of such simplification in Appendix \ref{appendix2}.

In real calculation, one can only include a finite number of hyperspherical harmonics, a truncation shall be made according to the symmetry properties of the system. Since we focus in this work on the tetraquark states with vanishing orbital angular momentum, the relevant hyperspherical harmonics are those corresponding to vanishing total orbital angular momentum $L$ and magnetic quantum number $M$, i.e. $L=M=0$. We choose all such hyperspherical harmonic functions with hyperangular quantum number $K\leq 3$. This leads to seven coupled differential equations which are numerically solved by using the inverse power method~\cite{H.W. Crater,SSZ}. The main advantage to take the inverse power method is its high precision for both ground and excited states. 

\section{Tetraquarks in vacuum}
\label{section3}

We start with computing the $cc\bar{c}\bar{c}$ and $bb\bar{b}\bar{b}$ bound states in vacuum. We employ the Cornell potential to describe the spin independent central interaction $V_{ij}^c$ between two quarks and the lattice result~\cite{Kawanai:2011jt} for the spin-spin coupling, 
\begin{eqnarray}
V^c_{ij}(|{\bf r}_{ij}|) &=& -{\alpha \over |{\bf r}_{ij}|}+\sigma |{\bf r}_{ij}|, \nonumber\\
V^{ss}_{ij}(|{\bf r}_{ij}|) &=& \beta e^{-\gamma |{\bf r}_{ij}| }.
\end{eqnarray}
The parameters $\alpha$, $\sigma$, $\beta$, $\gamma$ and the quark mass $m$ in the potential model are fixed by fitting the experimental data of charmonium and bottomonium masses. We calculate the quarkonium states $Q\bar Q$ via the two-body Schr\"odinger equation with the potential
\begin{equation}
V_{Q\bar Q} = {4\over 3}\left(V^c_{ij}(r) + V^{ss}_{ij}(r){\bf s}_i\cdot {\bf s}_j\right),  
\end{equation}
where the factor $4/3$ is the color factor for color-singlet states. With the model parameters presented in Table~\ref{tab1}, we obtain the quarkonium masses shown in Table~\ref{tab2}. One can see that, the potential model is effective in describing the heavy quarkonia. With the known parameters, we then solve the radial equations \eqref{radial1} and \eqref{radial2} for fully-heavy tetraquarks. It should be worth noting that similar calculations have been done in literatures~\cite{Debastiani:2017msn,Wang:2019rdo,Yang:2020rih,Chen:2020lgj,Liu:2019zuc}, by taking different potentials and employing different numerical method, e.g. variational method based on Gaussian expansion~\cite{Varga:1995dm,Suzuki:1998bn,SilvestreBrac:2007sg,SilvestreBrac:2008,Hiyama:2003cu}. 
\begin{table*}[!hbt]
\renewcommand\arraystretch{1.5}
\caption{Potential model parameters}
\label{tab1}
\setlength{\tabcolsep}{2.5mm}
\begin{tabular}{ccccccc}
	\toprule[1pt]\toprule[1pt] 
	$m_b$ & $m_c$ &
	$\alpha$ & $\sigma$ & $\gamma$ & $\beta_b$ & $\beta_c$\tabularnewline
	\midrule[0.7pt]
	4.7~GeV & 1.29~GeV &
	0.308 & 0.15~GeV$^2$ & 1.982~GeV & 0.239~GeV & 1.545~GeV
	\tabularnewline
	\bottomrule[1pt]\bottomrule[1pt]
	\end{tabular}
\end{table*}
\begin{table*}[!hbt]
\renewcommand\arraystretch{1.5}
\caption{The experimental~\cite{Zyla:2020zbs} and calculated quarkonium masses}
\label{tab2}
\setlength{\tabcolsep}{2.5mm}
\begin{tabular}{c|ccccccc}
	\toprule[1pt]\toprule[1pt] 
	State & $\eta_c$ & $J/\psi$ & $h_c(1P)$ & $\chi_c(1P)$ & $\eta_c(2S)$ & $\psi(2S)$ & $\chi_c(2P)$\tabularnewline
	\midrule[0.6pt]
	$M_E$(GeV) & 2.981& 3.097 & 3.525& 3.556 & 3.639 & 3.696 & 3.927 \tabularnewline
	$M_T$(GeV) & 2.968 &  3.102  & 3.480 & 3.500 & 3.654 & 3.720 & 4.000 \tabularnewline
	\midrule[1.2pt]
	State & $\eta_b$ & $\Upsilon(1S)$& $h_b(1P)$& $\chi_b(1P)$ & $\eta_b(2S)$ & $\Upsilon(2S)$ & $\chi_b(2P)$\tabularnewline
	\midrule[0.6pt]
	$M_E$(GeV) & 9.398 & 9.460 & 9.898 & 9.912 & 9.999 & 10.023 & 10.269 \tabularnewline
	$M_T$(GeV) & 9.397 & 9.459 & 9.845 & 9.860 & 9.957 & 9.977 &10.221 \tabularnewline
	\bottomrule[1pt]\bottomrule[1pt]
	\end{tabular}
\end{table*}

\begin{figure}[!tb]
\includegraphics[width=0.4\textwidth]{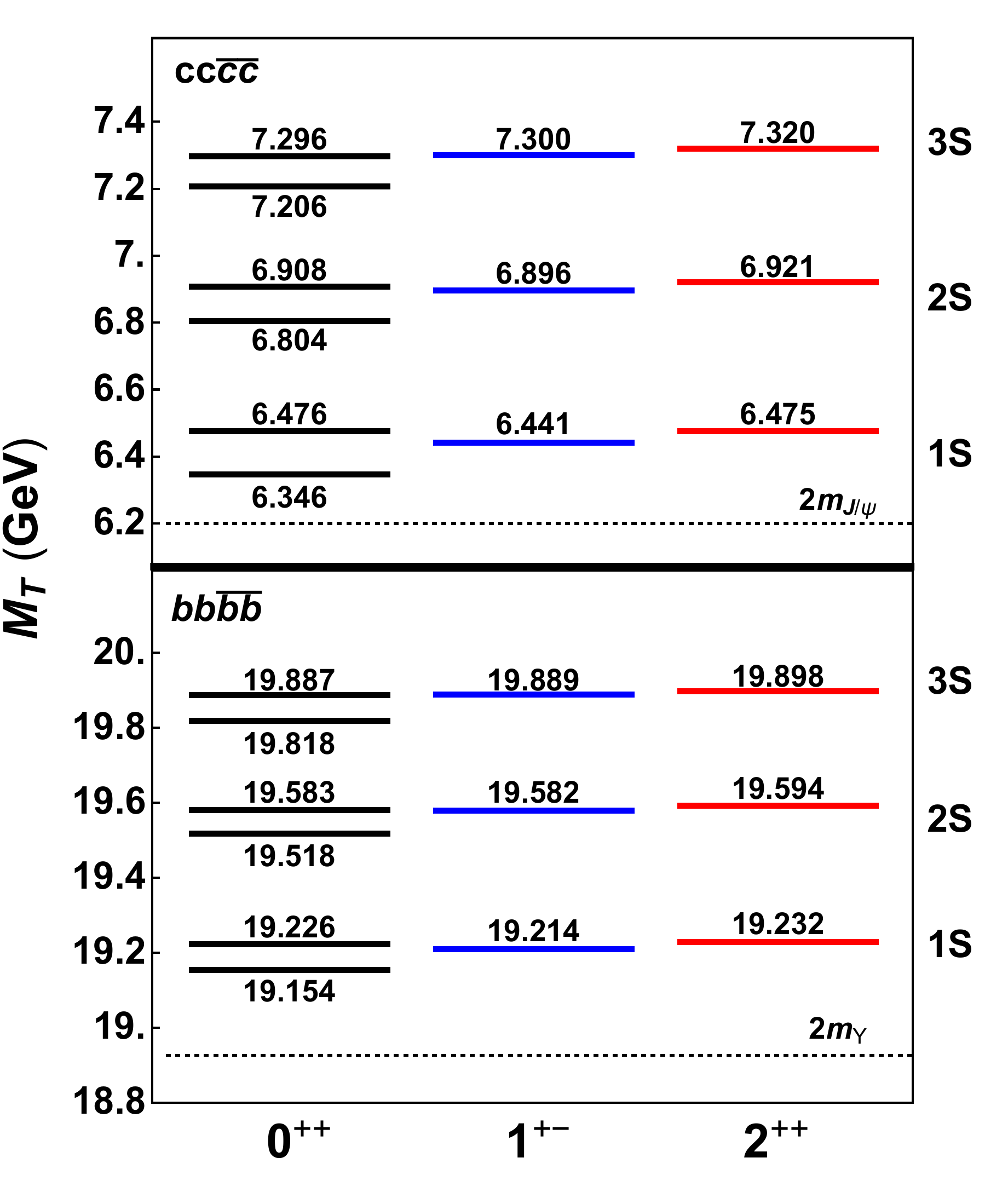}
\caption{The calculated tetraquark masses for the ground and radial-excited states, $1S,\ 2S$ and $3S$ of $cc\bar c\bar c$ (upper panel) and $bb\bar b\bar b$ (lower panel) with quantum numbers $J^{PC}=0^{++}$, $1^{+-}$ and $2^{++}$.}
\label{fig1}
\end{figure}

The tetraquark mass comes from the summation of the constituent masses $M=4m$ and the binding energy $E_r$ which is determined by the radial equations,
\begin{equation}
M_T=M+E_r \,.
\end{equation}
The root-mean-squared radius of the tetraquark state can be expressed as~\cite{Zhao:2020jqu} 
\begin{eqnarray}
r_{\text{rms}}^2&=&\Big\langle {1\over 4}\sum_{i=1}^4 ({\bf r}_i-{\bf X})^2\Big\rangle\nonumber\\
&=&\Big\langle{\mu\over 4m}({\bf x}_1^2+{\bf x}_2^2+{\bf x}_3^2)\Big\rangle \nonumber\\
&=&{\mu\over 4m} \int \sum_\kappa|R_\kappa(\rho)|^2 \rho^{10} d\rho.
\label{rms}
\end{eqnarray}
The root-mean-squared radius is $\mu$ independent. Taking $\mu=4m$ in our numerical calculations makes the prefactor be equal to unity, and the hyperradius $\rho$ can be considered as the radius of the tetraquark state. The calculated mass and mean radius for the ground and radial-excited states $1S$, $2S$ and $3S$ with quantum numbers $J^{PC}=0^{++},\ 1^{+-}$ and $2^{++}$ are shown in Table \ref{tab3}, where we have used the orthogonal and normalized condition for the hyper-spherical harmonic functions ${\mathcal Y}_\kappa(\Omega)$ and the normalization $\int \sum_\kappa|R_\kappa(\rho)|^2 \rho^8 d\rho=1$ for the radial functions $R_\kappa(\rho)$. The mass spectrum is also plotted in Figure \ref{fig1}. All the fully-heavy tetraquark states lay above the meson-meson mass threshold, $2m_{J/\psi}$ or $2m_{\Upsilon}$, shown as dotted lines in Figure \ref{fig1}. The $J^{PC}$ dependence of the mass is weak, and the small difference comes mainly from the spin-spin interaction. 
\begin{table*}[!tb]
\renewcommand\arraystretch{2.0}
\caption{The calculated tetraquark mass $M_T$ and the root-mean-squared radius $r_{\text{rms}}$ for the ground and radial-excited states, $1S,\ 2S$ and $3S$ of $cc\bar c\bar c$ and $bb\bar b\bar b$ with quantum numbers $J^{PC}=0^{++}$, $1^{+-}$, and $2^{++}$.}
\label{tab3}
 \setlength{\tabcolsep}{1.2mm}
\begin{tabular}{c|c|cc|cc|cc|ccc|ccc}
\toprule[1pt]\toprule[1pt] 
\multicolumn{2}{c|}{$J^{PC}$}&   \multicolumn{6}{c|}{$0^{++}$} &  \multicolumn{3}{c|}{$1^{+-}$} & \multicolumn{3}{c}{$2^{++}$}  \tabularnewline
\midrule[1pt]
\multicolumn{2}{c|}{State}&  \multicolumn{2}{c|}{$1S$} & \multicolumn{2}{c|}{$2S$} & \multicolumn{2}{c|}{$3S$} & $1S$ &  $2S$  & $3S$ & $1S$ &  $2S$ & $3S$\tabularnewline
\midrule[1pt]
\multirow{2}{*}{$cc\bar c\bar c$}
& $M_{T}$(GeV) & 6.346 & 6.476& 6.804 & 6.908 & 7.206  & 7.296 & 6.441& 6.896 & 7.300& 6.475 & 6.921 & 7.320 \tabularnewline
& $r_{\text{rms}}$(fm) & 0.323 & 0.351 & 0.445 & 0.457 & 0.550 & 0.530  &  0.331  & 0.446 & 0.547  &0.339 & 0.452 & 0.552 \tabularnewline
\midrule[1pt]
\multirow{2}{*}{$bb\bar b\bar b$}
 &$M_{T}$(GeV)& 19.154 & 19.226 & 19.518 & 19.583 &  19.818 & 19.887 & 19.214 & 19.582&  19.889  & 19.232 & 19.594 & 19.898 \tabularnewline
& $r_{\text{rms}}$(fm) & 0.180 & 0.186 & 0.259 & 0.259 & 0.328  & 0.325  &0.181  & 0.257 & 0.324  &0.183 & 0.259 & 0.326 \tabularnewline
\bottomrule[1pt]\bottomrule[1pt]
\end{tabular}
\end{table*}
\begin{table*}
	\renewcommand\arraystretch{2.0}
	\caption{The fraction of tetraquarks $cc\bar c\bar c$ and $bb\bar b\bar b$ with $J^{PC}=0^{++}$ in different color configures. }
	\label{tab4}
	\setlength{\tabcolsep}{1.0mm}
	\begin{tabular}{c|cc|cc|cc|cc|cc|cc}
		\toprule[1pt]\toprule[1pt] 
		\multicolumn{1}{c|}{} & \multicolumn{6}{c|}{$cc\bar c\bar c$} &   \multicolumn{6}{c}{$bb\bar b\bar b$} \tabularnewline
		\midrule[1pt]
		State & \multicolumn{2}{c|}{$1S$} & \multicolumn{2}{c|}{$2S$} &\multicolumn{2}{c|}{$3S$} & \multicolumn{2}{c|}{$1S$} & \multicolumn{2}{c|}{$2S$} & \multicolumn{2}{c}{$3S$}   \tabularnewline
		\midrule[1pt]
		$M_T$(GeV)& 6.346  & 6.476 & 6.804 & 6.908 & 7.206  &  7.296  & 19.154 & 19.226 & 19.518 & 19.583 & 19.818   & 19.887\tabularnewline
		\midrule[1pt]
		$|\phi_1\rangle$ &45.0\% & 54.2\% &29.8\% & 72.0\% & 19.9\%  & 65.6\%   &31.5\%  & 67.7\%  & 13.4\%  & 86.9\% &  6.2\%  & 94.1\%  \tabularnewline
		$|\phi_2\rangle$ &55.0\% & 45.8\% & 70.2\% & 28.0\% & 80.1\%   &   34.4\% &68.5\%  & 32.3\%  & 86.6\% & 13.1\%   &  93.8\%  &  5.9\%
		\tabularnewline
		\midrule[1pt]
		$|\phi_3\rangle$ &96.4\% & 6.3\% & 89.5\% & 21.2\% & 81.6\%  &  39.0\%  & 97.8\% & 3.6\% & 88.2\% & 16.6\% & 79.6\% & 25.8\% \tabularnewline
		$|\phi_5\rangle$ &3.6\%   & 93.7\% & 10.5\% & 78.8\% &  18.4\%  &  61.0\%  &2.2\% & 96.4\% & 11.7\% & 83.4\%   &  20.4\%  & 74.2\% \tabularnewline
		\midrule[1pt]
		$|\phi_4\rangle$ &6.8\% & 91.0\% & 23.9\% & 64.1\% & 38.5\%  &  50.6\%  &14.5\% & 84.6\% & 36.2\% & 58.8\%   &  49.6\%  & 44.8\%  \tabularnewline
		$|\phi_6\rangle$ &93.2\% & 9.0\% & 76.1\% & 35.9\% &  61.5\%  &  49.4\%  &85.5\% & 15.4\% & 63.8\% & 41.2\%   & 50.4\%   &  55.2\%
		\tabularnewline
		\bottomrule[1pt]\bottomrule[1pt]
	\end{tabular}
\end{table*}
\begin{figure}[!htb]
	\includegraphics[width=0.4\textwidth]{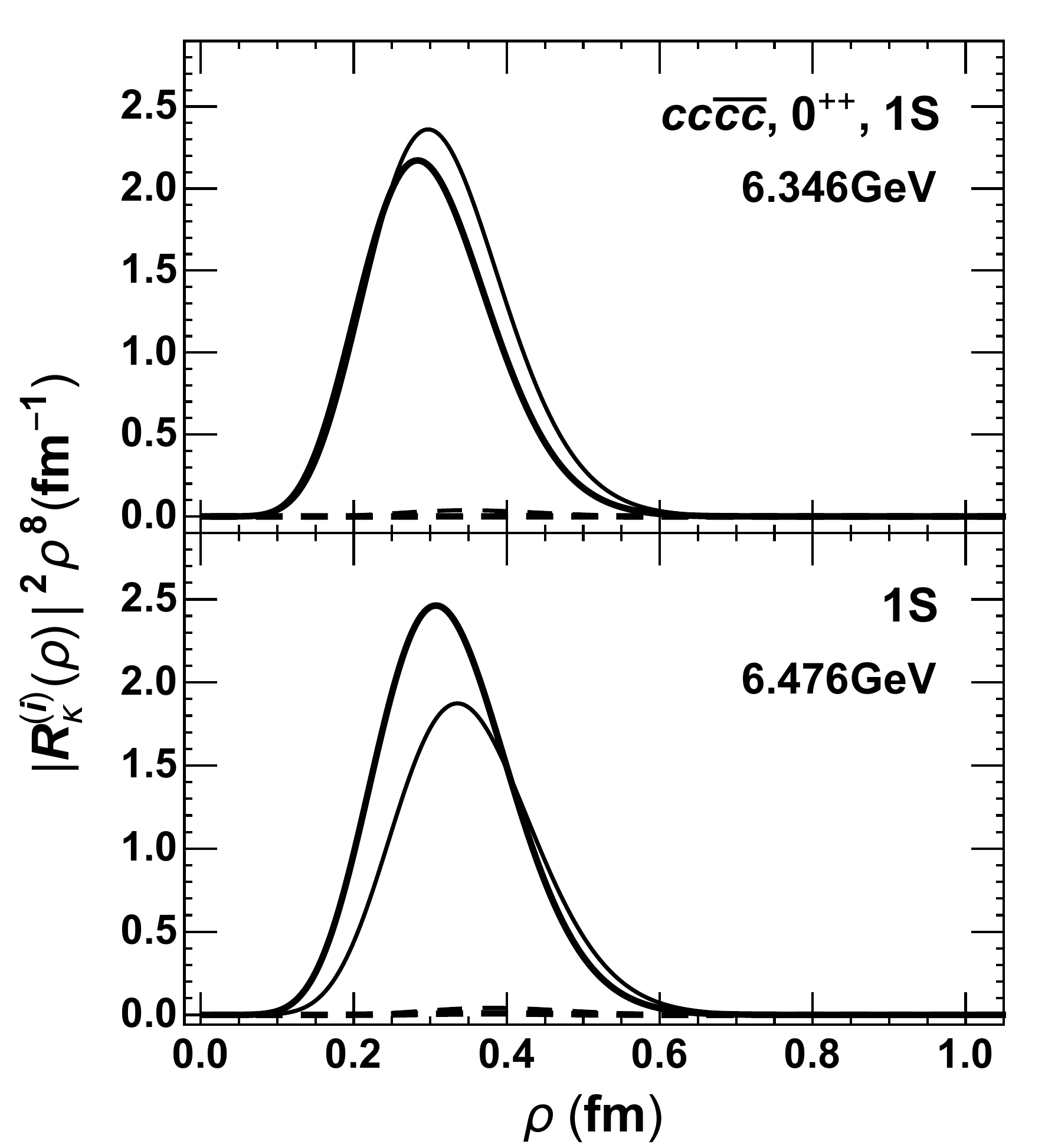}
	\caption{The radial probability fractions $|R_\kappa^{(i)}|^2\rho^8$ for the $cc\bar c\bar c$ $1s$ state with quantum number $J^{PC}=0^{++}$. The upper and lower panels correspond to the tetraquark mass $M_T=6.346$ and $6.476$ GeV, the thick and thin lines are the fractions $|R_\kappa^{(1)}|^2\rho^8$ and $|R_\kappa^{(2)}|^2\rho^8$, and the solid and dashed lines are with quantum number $\kappa=1$ and $2$. }
	\label{fig2}
\end{figure}

For the $0^{++}$ tetraquarks, there are two possible color-spin states $|\phi_1\chi_2\rangle$ and $|\phi_2\chi_1\rangle$ for any ground and radial-excited state. The mixture between the two color-spin states, see the coupling between the two radial functions $R^{(1)}$ and $R^{(2)}$ in \eqref{radial2}, will modify the tetraquark mass. The two modified masses are listed in Table \ref{tab3} and shown in Fig. \ref{fig1}. The left and right ones in Table \ref{fig3} and lower and higher ones in Fig. \ref{fig1} correspond to the modified results based on the states $|\phi_2\chi_1\rangle$ in the representation $6_c \otimes \bar 6_c$ and $|\phi_1\chi_2\rangle$ in $\bar 3_c \otimes 3_c$. 
To see the modification from the coupling between the two color-spin states, we show also the radial probability fractions in Figure \ref{fig2} for the ground state $1S$ of the fully-charmed tetraquark $cc\bar c\bar c$. The thick and thin lines represent the fractions $|R_\kappa^{(1)}|^2\rho^8$ and $|R_\kappa^{(2)}|^2\rho^8$. The small difference between the two shows a strong coupling, and the very small contributions from larger $\kappa$ indicate a very fast convergence in the numerical calculation. It is easy to understand that, for excited states the contributions from larger $\kappa$ should increase. To guarantee a good convergence for both ground and excited states, $\kappa$ runs from $1$ to $7$ in our calculation.   

A big problem in the study of multiquark states is to distinguish the multiquark states from molecular states. In the case of tetraquarks, while $|\phi_1\rangle$ and $|\phi_2\rangle$ form a complete and orthonormal set of color-singlet states, there can be other representations, especially the meson-meson states. For instance, a quark pair forms a meson state, and the other pair the other meson state, and then the two combine into a meson-meson state. Considering all the possible combinations, there are two meson-meson states,
\begin{eqnarray}
|\phi_3\rangle &=& |(Q_1\bar Q_3)_{1_c}(Q_2 \bar Q_4)_{1_c}\rangle,\nonumber\\
|\phi_4\rangle &=& |(Q_1\bar Q_4)_{1_c}(Q_2 \bar Q_3)_{1_c}\rangle.
\end{eqnarray}
These two states are not orthogonal to each other but orthogonal to one of the color-octet states 
\begin{eqnarray}
&& |\phi_5\rangle = |(Q_1\bar Q_3)_{8_c}(Q_2 \bar Q_4)_{8_c}\rangle,\nonumber\\
&& |\phi_6\rangle = |(Q_1\bar Q_4)_{8_c}(Q_2 \bar Q_3)_{8_c}\rangle
\end{eqnarray}
with
\begin{eqnarray}
&& \langle \phi_5|\phi_3\rangle = 0,\nonumber\\
&& \langle \phi_6|\phi_4\rangle = 0.
\end{eqnarray}
The above four molecular states can be expressed as a linear combination of the color-singlet states $|\phi_1\rangle$ and $|\phi_2\rangle$,
\begin{eqnarray} 
|\phi_3\rangle &=& \sqrt{1/3} |\phi_1\rangle + \sqrt{2/3} |\phi_2\rangle,\nonumber\\
|\phi_5\rangle &=& -\sqrt{2/3} |\phi_1\rangle + \sqrt{1/3} |\phi_2\rangle,
\end{eqnarray}
and
\begin{eqnarray}
|\phi_4\rangle &=& -\sqrt{1/3} |\phi_1\rangle + \sqrt{2/3} |\phi_2\rangle,\nonumber\\
|\phi_6\rangle &=& \sqrt{2/3} |\phi_1\rangle + \sqrt{1/3} |\phi_2\rangle.
\end{eqnarray}
Therefore, a tetraquark state with quantum number $J^{PC}=0^{++}$ can be expanded in color space in terms of either $|\phi_1\rangle$ and $|\phi_2\rangle$ or $|\phi_3\rangle$ and $|\phi_5\rangle$ or $|\phi_4\rangle$ and $|\phi_6\rangle$. The projection probabilities of each $0^{++}$ state are shown in Table \ref{tab4}. 

Finally, we look at the exotic hadron $X(6900)$ recently observed by the LHCb Collaboration~\cite{Aaij:2020fnh}. The current experiment measures only the mass and the width, and is not able to determine the spin and parities $J^{PC}$. Our theoretical result indicates that, $X(6900)$ maybe the first radial excited state $2S$ of $cc\bar c\bar c$ with $J^{PC}=0^{++}$ (6908 MeV) or $1^{+-}$ (6896 MeV). No matter what the quantum number $J^{PC}$ is, $X(6900)$ is likely to be a tetraquark state, instead of a meson-meson state.   

\section{Tetraquarks in hot medium}
\label{section4}
It is widely accepted that there exists a deconfinement phase transition from hadron gas to quark matter at high temperature and baryon density. From the lattice QCD simulations~\cite{Bazavov:2011nk,Fodor:2004nz} and many effective model studies~\cite{Xin:2014ela,Li:2018ygx}, the critical temperature of the transition is about $T_c = 165$ MeV at zero baryon density.
Considering the fact that heavy-quark mass is much larger than the temperature scale, the tightly bound states of heavy quarks, such as $J/\psi$ and $\Upsilon$, can survive in the quark matter and be used to probe the properties of the new matter. In this section, we study the temperature dependence of the tetraquark properties and their dissociation temperatures in the QGP.
 
In the color-deconfined QCD medium, the heavy-quark potential is screened, and the long-range interaction is strongly weakened when the temperature is high enough. The lattice QCD simulations indicate that, the finite-temperature potential between a pair of heavy quarks can be approximated by the free energy $F(r,T)$~\cite{Petreczky:2010yn,Lafferty:2019jpr,Zhao:2020jqu}. For the heavy-quark bound-states in the hot QCD medium, we take the free energy $F(r,T)$ as the central potential $V_{ij}^c(|{\bf r}_{ij}|,T)$, and neglect the finite-temperature corrections to the spin-spin interaction,
\begin{eqnarray}
V^c_{ij} &=& {1\over \Gamma(3/4)}{\sigma\over \mu}\left[{\Gamma(1/4)\over 2^{3/2}}-{\sqrt{\mu|{\bf r}_{ij}|}\over 2^{3/4}}K_{1/4}\left(\mu^2|{\bf r}_{ij}|^2\right)\right]\nonumber\\
&-&\alpha\left[\mu+{e^{-\mu|{\bf r}_{ij}|}\over |{\bf r}_{ij}|}\right],
\label{vt} 
\end{eqnarray}
where $\Gamma$ and $K_{1/4}$ are the Gamma functions and modified Bessel function of the second kind. The temperature dependent screening mass $\mu(T)$ is extracted from fitting the lattice data. 

\begin{figure}[!htb]
	$$\includegraphics[width=0.4\textwidth]{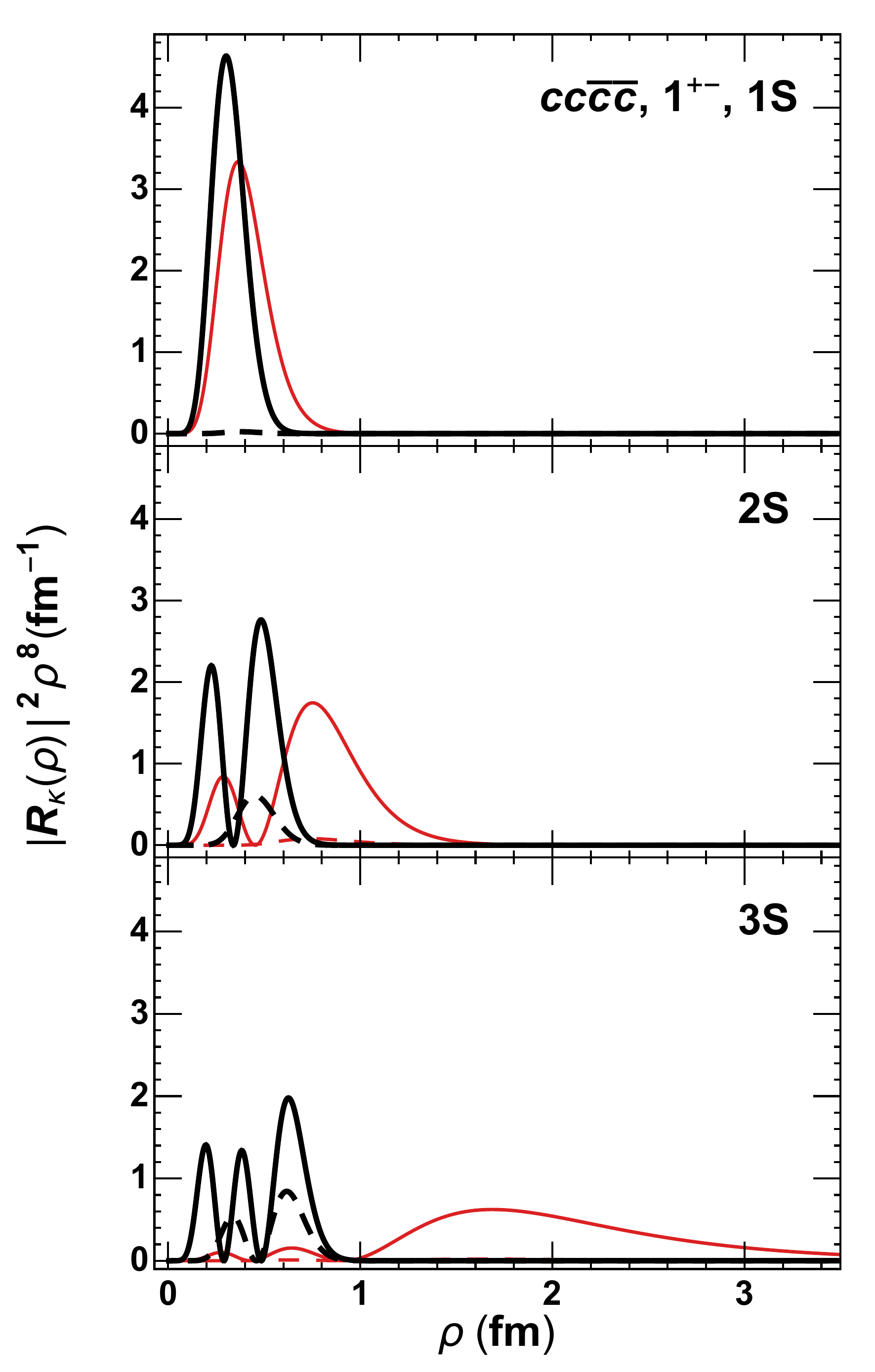}$$ 
	\caption{The radial probabilities for the ground and radial excited states, $1S,\ 2S$ and $3S$ of the tetraquark $cc\bar c\bar c$ with quantum number $J^{PC}=1^{+-}$ in vacuum (thick lines) and at critical temperature $T_c$ (thin lines). The solid and dashed lines are with quantum number $\kappa=1$ and $2$. }
	\label{fig3}
\end{figure}

We solve again the coupled radial equations \eqref{radial1} and \eqref{radial2} with the central interaction \eqref{vt} and obtain the binding energy $E_r(T)$ and relative wave function $\Phi(\rho,\Omega,T)$ as functions of temperature. The radial probabilities for the ground and excited states of the fully-charmed tetraquark $cc\bar c \bar c$ with $J^{PC}=1^{+-}$ at the critical temperature and the comparison with the vacuum result are shown in Fig. \ref{fig3}. At finite temperature, the long range confinement force is suppressed and the interaction strength is weakened due to the color screening. As a result, the wave function expands outward, and the averaged size of the tetraquark becomes larger in comparison with vacuum, especially for the excited states. The temperature effect changes also the radial symmetry of the system. The random thermal motion of the heavy quarks will smear the angle dependence of the wave function, and the asymmetric components with larger values of the hyperangular quantum number $\kappa$ are suppressed. These features for tetraquarks are consistent with the properties of quarkonia and heavy-flavor baryons at finite temperature~\cite{Guo:2012hx,Shi:2019tji}. 

From Fig. \ref{fig3} the wave function for the second radial excited state $3s$ expands with temperature very fast, and the peaks in vacuum almost disappear at the critical temperature $T_c$. This means that the bound state is close to the disappearance. Similar to the definition for quarkonium dissociation, the tetraquark dissociation temperature $T_d$ is defined as the divergence of the size and the vanish of the binding energy,
\begin{eqnarray}
\langle \rho\rangle (T_d) &\to& \infty,\nonumber\\
E_r(T_d) &\to& 0.
\end{eqnarray}
The dissociation temperatures for different tetraquark states are shown in Table \ref{tab5}. Considering the very weak $J^{PC}$ dependence of the tetraquark mass $M_T$ shown in Fig. \ref{fig1}, the dissociation temperature is almost independent of the quantum numbers $J^{PC}$, and we have neglected this small difference in Table \ref{tab5}. Different from the $cc\bar c\bar c$ states which are already dissociated a little bit above the critical temperature, the $bb\bar b\bar b$ states can survive in the QGP phase at very high temperature, due to the extremely large mass of $b$ quark. It is clear that, the excited states will disappear first. 
\begin{table}[!htb]
	\renewcommand\arraystretch{1.5}
	\caption{The scaled dissociation temperatures $T_d$ for fully-heavy tetraquarks. $T_c$ is the critical temperature of the deconfinement phase transition. }
	\label{tab5}
	\setlength{\tabcolsep}{1.5mm}
	\begin{tabular}{c|ccc|ccc}
		\toprule[1pt]\toprule[1pt] 
		& \multicolumn{3}{c|}{$cc\bar c\bar c$} & \multicolumn{3}{c}{$bb\bar b\bar b$}\tabularnewline
		\cline{2-7}
		& $ 1S $ & $2S$& $3S$&  $1S$ & $2S$& $3S$ \tabularnewline
		\midrule[1pt]
		$T_d / T_c$ & 1.08 & 1.02 & 1.0 & 2.40 & 1.85 & 1.30 
		\tabularnewline
		\bottomrule[1pt]\bottomrule[1pt]
	\end{tabular}
\end{table}

\section{Tetraquark production in nuclear collisions}
\label{section5}
The deconfinement phase transition can be realized in the early stage of relativistic heavy-ion collisions at RHIC and LHC when the temperature of the system is above the critical temperature $T_c$. The appearance of QGP significantly changes the production mechanism of hadrons. In particular, the production of low-momentum hadrons are dominantly contributed by the coalescence of partons when the QGP cools down due to the expansion of the colliding system and the temperature reaches $T_c$. The coalescence model~\cite{Fries:2003kq} has successfully described the light hadron production in heavy-ion collisions, especially the quark number scaling of the elliptic flow~\cite{Molnar:2003ff,Lin:2002rw} and the enhancement of the baryon to meson ratio~\cite{Hwa:2002tu,Oh:2009zj}. Since heavy quarks are rare particles in the QGP, their hadronization is more in line with the spirit of the coalescence mechanism. The production of quarkonia and multi-charmed baryons in heavy-ion collisions are well studied in different coalescence models~\cite{Greco:2003vf,Zhao:2016ccp,He:2014tga}. It shows that their production in heavy-ion collisions is largely enhanced due to the combination of uncorrelated charm quarks in the QGP~\cite{Zhao:2017gpq}. This provides a most probable way to discover those multi-charmed baryons like $\Xi_{cc}$ and $\Omega_{ccc}$ in heavy-ion collisions at the RHIC and LHC energies. On the other hand, the previous studies on exotic hadron production in heavy-ion collisions show that, the yields of exotic hadrons are expected to be strongly affected by their structures~\cite{Cho:2017dcy,Fontoura:2019opw}. Considering the fact that, bottom quarks are rarely created even in heavy-ion collisions at LHC energy, we discuss only the production of fully-charmed tetraquark $cc\bar c\bar c$ in this section. Taking into account the $cc\bar c\bar c$ dissociation temperature which is almost the same as the critical temperature, the initially produced tetraquarks via nucleon-nucleon collisions will be dissociated in the QGP phase and all the tetraquarks measured in the final state are from the coalescence at the freeze-out of the QGP.     

In the coalescence model, the differential production cross section of a tetraquark state is given by
\begin{eqnarray}
{d\sigma \over d^2{\bf P}_T dy} &=& C \sigma_{NN}^{inel} n_{Q Q \bar Q \bar Q}^{AA}\int_\Sigma {P^\mu d\sigma_\mu(R) \over (2\pi)^3}\int{d^9{\bf x}d^9{\bf p} \over (2\pi)^9}\nonumber\\
&\times& f_{Q_1} f_{Q_2} f_{\bar Q_3} f_{\bar Q_4} W({\bf x},{\bf p}),
\label{coalescence}
\end{eqnarray}
where  $R_\mu =\sum_i (t_i, {\bf r}_i)/4$ is the four dimensional center-of-mass coordinate of the tetraquark, and $P_\mu$ represents its four-momentum with $P_0 =\sqrt{M_T^2+{\bf P}^2}$ being the energy, ${\bf P}=\sum_i {\bf q}_i$ the total three-momentum, ${\bf P}_T$ the transverse-momentum, and $y$ the rapidity. The nine-dimensional coordinate and momentum ${\bf x}$ and ${\bf p}$ are the shorthands of three relative coordinates and momenta ${\bf x}_i$ and ${\bf p}_i\ (i=1,2,3)$. They are defined in the rest-frame of the tetraquark. The factor $C=1/1296,\ 1/432$ and $5/1296$ for $J^{PC}=0^{++},\ 1^{+-}$ and $2^{++}$ states are from the statistics determined by the intrinsic symmetry, i.e. color, spin, and isospin, and $\sigma_{NN}^{inel}$ is the inelastic cross section of the corresponding nucleon-nucleon collisions.

In heavy-ion collisions at RHIC and LHC energies, the heavy quarks in the QGP phase are almost all created through the initial nucleon-nucleon collisions. For the four heavy quarks $Q,\ Q,\ \bar Q$ and $\bar Q$ to form a tetraquark state, they can be from two, or three or four binary collisions. In a heavy-ion collision ($AA$) with fixed number $N_{\text{coll}}$ of binary nucleon-nucleon ($NN$) collisions, the averaged number of combinations to have four quarks reads 
\begin{eqnarray}
\label{n4q}
n_{Q Q \bar Q \bar Q}^{AA} &=& 2N_{\text{coll}}(N_{\text{coll}}-1)(n_{Q\bar Q}^{NN})^2\\
                           &+& 4N_{\text{coll}}(N_{\text{coll}}-1)(N_{\text{coll}}-2)(n_{Q\bar Q}^{NN})^3\nonumber\\
                           &+& N_{\text{coll}}(N_{\text{coll}}-1)(N_{\text{coll}}-2)(N_{\text{coll}}-3)(n_{Q\bar Q}^{NN})^4\nonumber,
\end{eqnarray}
where $n_{Q\bar Q}^{NN}$ represents the averaged pair number of heavy quarks created in a nucleon-nucleon collision. At the colliding energy $\sqrt s=2.76$ TeV, the inelastic cross section is $\sigma_{NN}^{inel}=65$ mb~\cite{Abelev:2012sea}, and the production cross sections of heavy quarks are $d\sigma_{c\bar c} / dy=0.7$~mb~\cite{Abelev:2012vra} and $d\sigma_{b\bar b}/dy=15\ \mu$b~\cite{Cacciari:2012ny}. Therefore we have the averaged number $n_{c\bar c}^{NN} = (d\sigma_{c\bar c} / dy)/\sigma_{NN}^{inel} = 1.1\times 10^{-2}$ and $n_{b\bar b}^{NN} = (d\sigma_{b\bar b}/dy)/\sigma_{NN}^{inel} = 2.3\times 10^{-4}$. In the calculation we have neglected the probability of producing two pairs of heavy quarks in a nucleon-nucleon collision.

The integration region $\Sigma$ in the coalescence model \eqref{coalescence} is the isothermal hadronization hypersurface controlled by the critical temperature $T(R_\mu \in \Sigma) = T_c$, and the integration element $d\sigma_\mu$ is the normal four-vector to $\Sigma$. Such an isothermal hypersurface can be extracted from hydrodynamic calculations. 

The space-time evolution of the QGP phase can be successfully described by relativistic hydrodynamics~\cite{Kolb:2003dz}. The theory is based on the conservation laws of the matter. For ideal hydrodynamics without considering the dissipation of the fluid, the evolution of the QGP is governed by the energy-momentum conservation and baryon number conservation, 
\begin{eqnarray}
\label{hydro}
&& \partial_\mu T^{\mu\nu}=0,\nonumber\\
&& \partial_\mu n^{\mu}=0, 
\end{eqnarray}
where $T^{\mu\nu}=(\epsilon + P)u^\mu u^\nu - Pg^{\mu\nu}$ is the energy-momentum tensor with $\epsilon$ being the energy density, $P$ the pressure and $u^\mu$ the fluid velocity, and $n_\mu=n u_\mu$ is the baryon current with $n$ being the baryon number density. $\epsilon,\ P$ and $n$ are functions of temperature $T$ and baryon number $n$, given by the equation of state of the hot medium. To compute the equation of state, we treat the deconfined phase at high temperature as an ideal gas of gluons, massless $u$ and $d$ quarks, as well as $s$ quarks with mass $m_s = 150$ MeV, and the hadron phase at low temperature as an ideal gas of all known hadrons and resonances with mass up to 2 GeV~\cite{Sollfrank:1996hd}. The phase transition temperature is chosen as $T_c=165$ MeV. The initial condition of the Eq. \eqref{hydro} at proper time $\tau_0=0.6$ fm/c is determined by the colliding energy and nuclear geometry, which lead to a maximum initial temperature $T_0=484$ MeV for central Pb-Pb collisions at $\sqrt{s_{NN}}=2.76$ TeV. For such extremely high-energy nuclear collisions, the baryon number density approaches to zero. By solving the hydrodynamic equations, we obtain the space-time profiles of temperature $T(t,{\bf r})$ and fluid velocity $u^\mu(t,{\bf r})$. Based on the hydrodynamic profiles, we can determine the isothermal hadronization hypersurface $\Sigma$ and its normal four-vector $d\sigma_\mu$ at the hadronization temperature $T_c$. 
 
There are two key ingredients in the coalescence model \eqref{coalescence}. One is the phase space distribution of heavy quarks $f_Q(t,{\bf r},{\bf q})$ and $f_{\bar Q}(t,{\bf r},{\bf q})$, and the other is the coalescence probability $W({\bf x},{\bf p})$ (Wigner function) for four heavy quarks to combine into a tetraquark. We first consider $f_Q$ and $f_{\bar Q}$. Since fully-bottomed tetraquarks are extremely rarely produced in heavy-ion collisions at RHIC and LHC energies, we will only calculate here charmed tetraquarks. In Pb-Pb collisions at energy $\sqrt {s_{NN}}=2.76$ TeV, the experimental measurement on $D$-meson elliptic flow~\cite{Abelev:2013lca} indicates that, charm quarks reaches kinetic equilibrium with the QGP. Therefore, we can take the normalized Fermi-Dirac distribution $f_{FD}(q)=A/(e^{u_\mu q^\mu/T}+1)$ as the charm quark and anti-quark momentum distribution, where $u_\mu(t,{\bf r})$ is the local fluid velocity of the matter determined by the hydrodynamics \eqref{hydro}, and $A(t,{\bf r})$ is the normalization factor. The number density $n_c(t,{\bf r})$ in coordinate space is controlled by the charm conservation law for the charm current $n_c^\mu=n_c u^\mu$,
\begin{equation}
\partial_\mu n_c^\mu=0
\end{equation}      
with the initial condition
\begin{equation}
n_c(t_0,{\bf r}) = {\sigma_{NN}^{inel} \cosh y \over N_{\text{coll}}\tau_0}T_A\left({\bf r}_T+{{\bf b}\over 2}\right)T_B\left({\bf r}_T-{{\bf b}\over 2}\right),
\end{equation}
where $T_A({\bf r}_T+{\bf b}/2)$ and $T_B({\bf r}_T-{\bf b}/2)$ are thickness functions of the two colliding nuclei, ${\bf r}_T$ is the transverse coordinate, and ${\bf b}$ the impact parameter of the collision. Combining the momentum and spatial distributions, we obtain the phase-space distribution function for charm quarks
\begin{equation}
f_c(t,{\bf r}, {\bf q}) = n_c(t, {\bf r})f_{FD}(t,{\bf r},{\bf q}).
\end{equation}

We now come to the Wigner function which reflects the dynamics of the hadronization of heavy quarks in hot medium. For light hadrons and light-heavy systems, the nonperturbative (confinement) properties of hadronization makes it difficult to theoretically calculate the Wigner function. It is usually to take a double Gaussian distribution~\cite{Fries:2003kq,Molnar:2003ff,Lin:2002rw,Hwa:2002tu,Oh:2009zj} in the phase space with adjustable parameters which can be fixed by fitting the experimental data. For fully-heavy tetraquark systems, however, one can nonperturbatively solve the corresponding Schr\"odinger equation with confinement potential and obtain the wave function $\Phi({\bf x},T)$ of the system and in turn the Wigner function via a Fourier transformation,
\begin{equation}
W({\bf x},{\bf p},T)=\int d^9{\bf y}e^{-i{\bf p}\cdot {\bf y}}\Phi\left({\bf x}+{{\bf y}\over 2},T\right)\Phi\left({\bf x}-{{\bf y}\over 2},T\right).
\label{wigner}
\end{equation}
Note that, the Schr\"odinger equation is solved in the QGP phase, the medium properties are reflected in the Wigner function. 
\begin{figure}[!htb]
\includegraphics[width=0.4\textwidth]{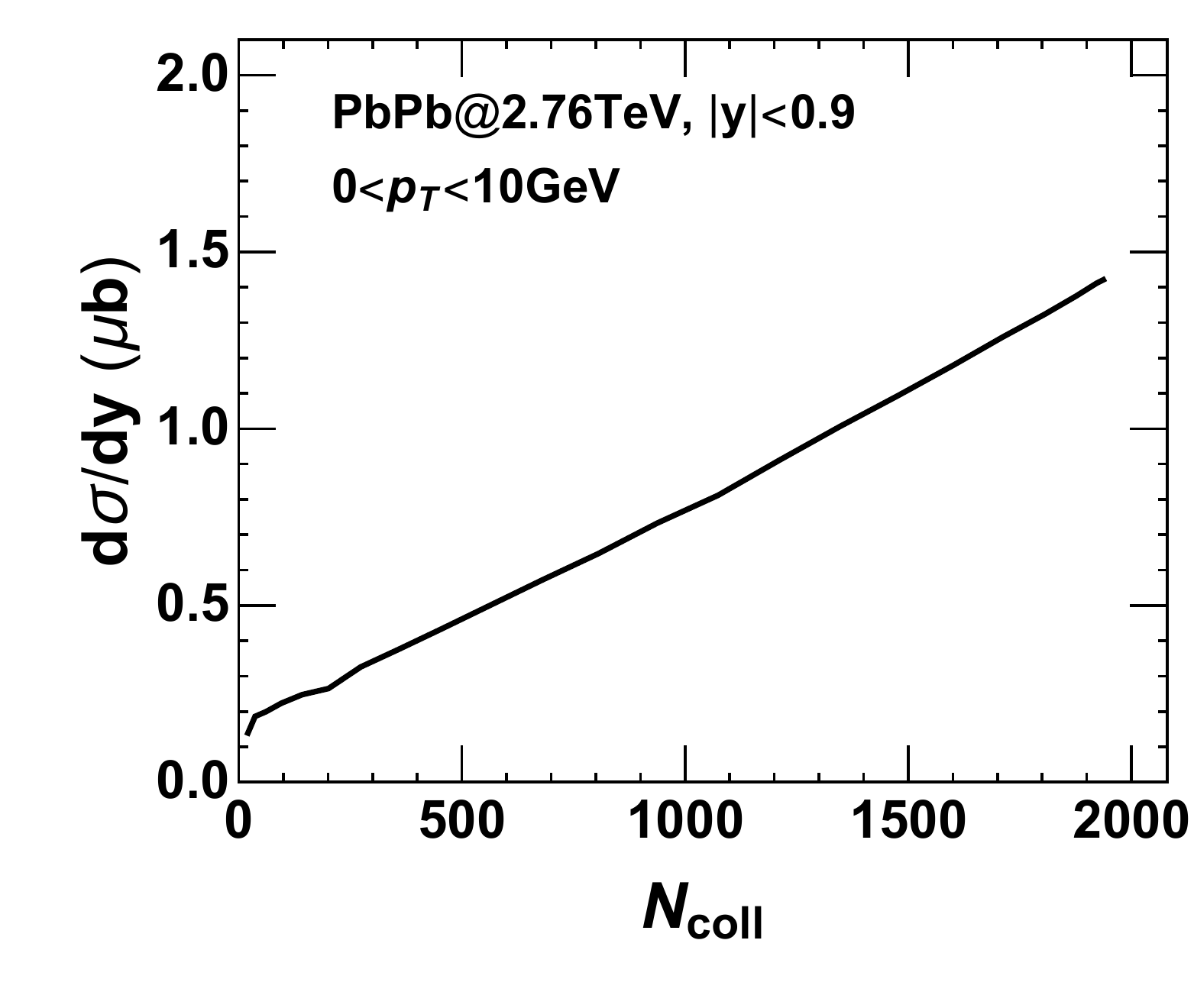}
\caption{ The tetraquark $cc\bar c\bar c$ production yield per unit rapidity as a function of the number of binary collisions in Pb-Pb collisions at LHC energy. }
\label{fig4}
\end{figure}
\begin{figure}[!htb]
	\includegraphics[width=0.4\textwidth]{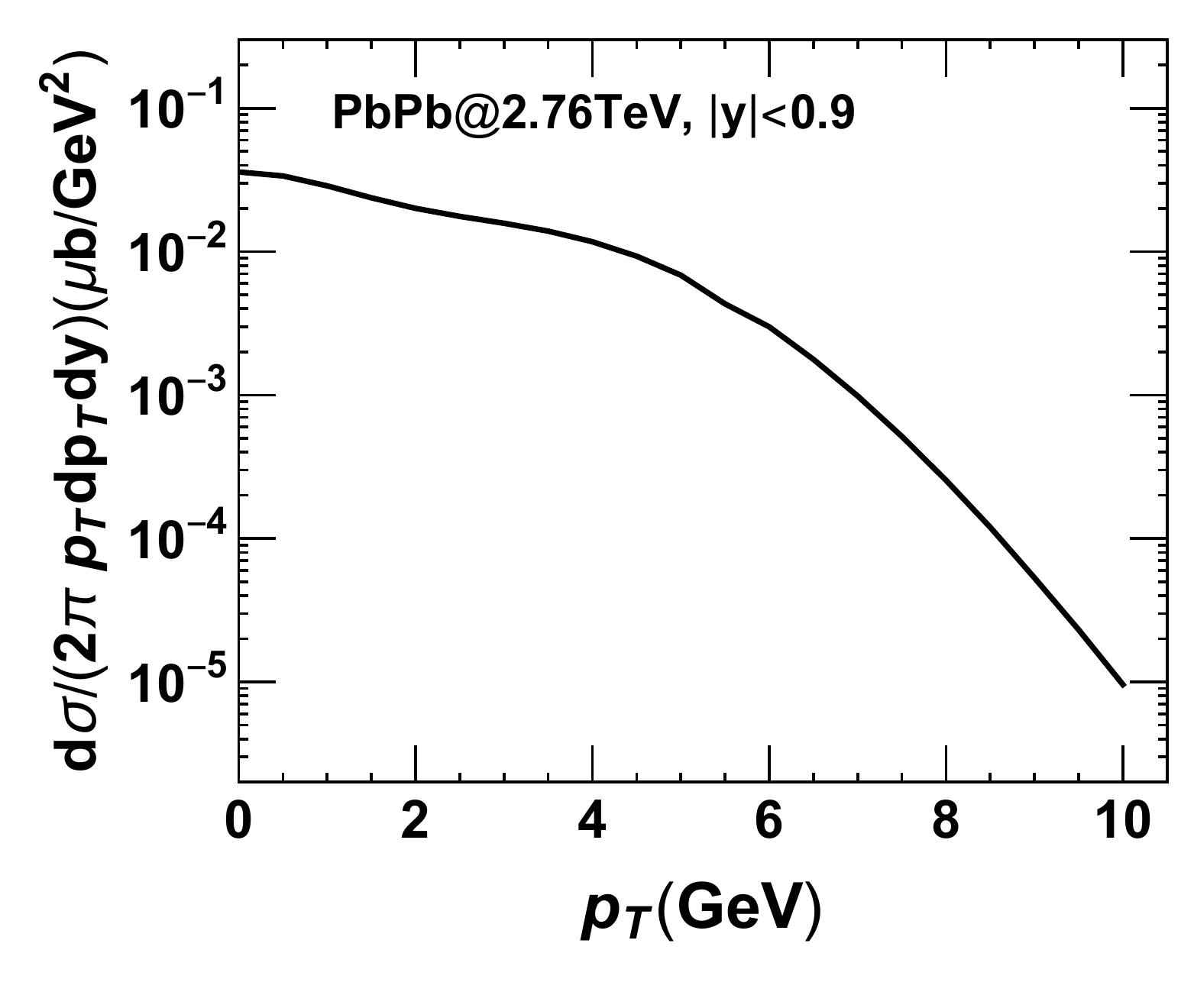}
	\caption{ The tetraquark $cc\bar c\bar c$ transverse momentum distribution in most central Pb-Pb collisions at LHC energy. }
	\label{fig5}
\end{figure}

Taking all the ingredients discussed above for the coalescence model \eqref{coalescence}, we calculated numerically the fully-charmed tetraquark $cc\bar c\bar c$ yield and transverse momentum distribution in Pb-Pb collisions at LHC energy. The results are shown in Figs. \ref{fig4} and \ref{fig5}. Here we have included all the tetraquark states $1S,\ 2S$ and $3S$ with $J^{PC}=0^{++},\ 1^{+-}$ and $2^{+-}$. 

It is easy to understand the strong tetraquark enhancement in heavy-ion collisions in comparison with nucleon-nucleon collisions, because a nuclear collision consists of $N_{\text{coll}}$ nucleon-nucleon collisions. In a most central Pb-Pb collision, $N_{\text{coll}}$ reaches 1937. Let us consider the tetraquark yield in a binary nucleon-nucleon collision. From the previous study~\cite{Karliner:2016zzc,Aaij:2011yc,Khachatryan:2016ydm}, the tetraquark production cross section in p-p collisions is $d\sigma_{pp}^{cc\bar c\bar c}/dy = 78$~pb at $\sqrt s=7$ TeV. In a most central Pb-Pb collision the effective cross section per binary collision is $d\sigma_{AA}^{cc\bar c\bar c}/dy/N_{\text{coll}} = 0.77$~nb, which is almost $10$ times larger than that in corresponding p-p collisions. The reason for this nontrivial enhancement is from the many combinations for having four quarks to form a tetraquark state. It is highly nonlinear in $N_{\text{coll}}$, see \eqref{n4q}.  

The difference between nucleus-nucleus and nucleon-nucleon collisions is not only the yield but also the momentum distribution. In nucleon-nucleon collisions, the initially created heavy quarks via hard processes carry high momentum, and the produced tetraquarks will inherit the high momentum. In nucleus-nucleus collisions, the heavy quarks lose energy when they pass through the medium and get thermalized before the hadronization. Therefore, the formed tetraquarks via coalescence mechanism are mainly distributed in low momentum region, see Fig. \ref{fig5}.   

\section{Summary}
\label{section6}
In this paper, we solved the four-body Schr\"odinger equation and investigated the properties of fully-heavy tetraquark states $cc\bar c\bar c$ and $bb\bar b\bar b$ in vacuum and at finite temperature.
To increase the precision of solving the equation, we expanded the wave functions in series of hyperspherical harmonics and obtained the eigenstates and eigenvalues by using an iteration algorithm based on the inverse power method. This algorithm allows us to study not only the ground but also excited tetraquark states.  

In vacuum, we found that the masses of all the tetraquark states $1S,\ 2S$ and $3S$ with $J^{PC}=0^{++},\ 1^{+-}$ and $2^{++}$ are above the $2m_{J/\psi}$ or $2m_\Upsilon$ threshold. The experimentally observed exotic state $X(6900)$ is likely to be a tetraquark state of $cc\bar c\bar c$,  and the possible quantum number is $J^{PC} = 0^{++}$ or $1^{+-}$. 

At finite temperature, we determined the tetraquark dissociation temperatures due to the color screening effect on the heavy-quark potential. $bb\bar b\bar b$ can survive in almost all the QGP phase, while $cc\bar c\bar c$ is already melted at the critical temperature $T_c$. Taking the wave function at finite temperature, we constructed, without introducing any adjustable parameter, the Wigner function in phase space which is the key ingredient of the coalescence mechanism. In the framework of coalescence model, we calculated the production cross section and transverse momentum distribution for $cc\bar c\bar c$ in heavy-ion collisions. Compared to p-p collisions, the production yield, not only for $A$-$A$ but also for a binary collision, is extremely enhanced in heavy-ion collisions, and the tetraquarks are mainly distributed at low momentum. 

Due to the complicated background in nuclear collisions, it is challenging to search for rare particles with low/median $p_T$ in heavy-ion collisions. However, for fully-heavy tetraquarks, the four-lepton decay channel $X\to l_1^+l_2^-l_3^+l_4^-$ can be well separated from the bulk background and makes it possible to find such exotic states in low $p_T$ region~\cite{Karliner:2016zzc}. In central collisions, the production cross section of fully-charmed tetraquarks is around three or four orders of magnitude larger than that in p-p collisions, and the leptons produced in the decay channel are energetic but do not interact with the hot medium. Consequently, we expect that the fully-charmed tetraquark shall be able to be measured by lepton detectors at LHC. This calls for theoretical predictions of the cross section for the four-lepton decay~\cite{Becchi:2020mjz,Becchi:2020uvq}, and a systematic study of both the total and differential cross sections for all possible fully-heavy tetraquarks is required. The results will be reported in our future work.

{\bf Acknowledgement:} We thank Guojun Huang and Lu Meng for helpful discussions during the work. This work is supported by the NSFC Grant Number 11890712, and S.S. is grateful to the Natural Sciences and Engineering Research Council of Canada.

\begin{appendix}
\section{Hyperspherical harmonic functions}
\label{appendix1}

For a four-body system with central two-body interaction, the conserved quantities include the orbital angular momenta $\hat{\bm l}_1,\ \hat{\bm l}_2$ and $\hat{\bm l}_3$ corresponding to the relative coordinates ${\bf x}_1,\ {\bf x}_2$ and ${\bf x}_3$ and $\widehat{\bf L}_1=\hat{\bm l}_1,\ \widehat{\bf L}_2 = \hat{\bm l}_1+ \hat{\bm l}_2$ and $\widehat{\bf L} = \widehat{\bf L}_3 = \hat{\bm l}_1 + \hat{\bm l}_2 + \hat{\bm l}_3$ for the 1-2 sub-system, 1-2-3 sub-system and whole four-body system, and the projections $\widehat L_{1z},\ \widehat L_{2z}$ and $\widehat L_{3z}$. Any two of these operators are commutative and their eigenvalues, $L_2, L, M_2, M, l_1, l_2, l_3, n_2$ and $n_3$, form a complete set of quantum numbers.

The hyperspherical harmonic functions ${\mathcal Y}_\kappa(\Omega)$ are defined as the eigenstates of the hyperangular momentum $\widehat {\bf K}_3^2$ of the system,
\begin{eqnarray}
&& \widehat {\bf K}_3^2 {\mathcal Y}_{\kappa}(\Omega) = K(K+7) {\mathcal Y}_{\kappa}(\Omega),\\
&& \widehat {\bf K}_3^2 = -{\partial^2 \over \partial \alpha_3^2}+{3-7\cos(2\alpha_3) \over \sin(2\alpha_3)}{\partial \over \partial \alpha_3}+{1 \over \cos^2 \alpha_3} \widehat {\bf K}_2^2\nonumber\\
&& \qquad+{1\over \sin^2 \alpha_3}\hat {\bm l}^2_3,\nonumber\\
&& \widehat {\bf K}_2^2 = -{\partial^2 \over \partial \alpha_2^2}-{4\cos(2\alpha_2) \over \sin(2\alpha_2)}{\partial \over \partial \alpha_2}+{1\over \cos^2 \alpha_2} \hat {\bm l}^2_1 +{1\over \sin^2 \alpha_2}\hat {\bm l}^2_2\nonumber
\end{eqnarray}
with the solution
\begin{eqnarray}
K &=& 2(n_2+n_3)+l_1+l_2+l_3,\\
{\mathcal Y}_{\kappa} &=&\prod_{i=2}^3N_i(\sin \alpha_i)^{l_i}(\cos\alpha_i)^{K_{i-1}}P_{n_il_iK_i}(\cos 2\alpha_i)\nonumber\\
&\times&\sum_{m_1,m_2,m_3}\prod_{j=2}^3\langle L_{j-1}M_{j-1}l_jm_j|L_jM_j\rangle  \nonumber\\
&\times& \prod_{k=1}^3Y_{l_k,m_k}(\theta_k,\phi_k),\nonumber
\end{eqnarray}
where $N_i$ is the normalization coefficient
\begin{equation}
N_i=\sqrt{(2K_i+4)n_i! \Gamma(n_i+K_{i-1}+l_i+2) \over \Gamma(n_i+l_i+{3\over 2}) \Gamma(n_i+K_{i-1}+{3\over 2})},
\end{equation}
$P_{n_il_iK_i}\equiv P_{n_i}^{l_i+1/2,K_{i-1}+(3j-5)/2}$ is the Jacobi polynomial, $Y_{l_k,m_k}(\theta_k,\phi_k)$ are the ordinary spherical harmonic functions, and $\kappa$ stands for all the quantum numbers. 

Considering only the radial excited states with quantum numbers $L=M=0$, we explicitly list here the first seven hyperspherical harmonic functions ${\mathcal Y}_\kappa(\Omega)$ with quantum numbers $(\kappa, K, n_3, n_2, l_1, l_2, l_3)=$ (1, 0, 0, 0, 0, 0, 0), (2, 2, 1, 0, 0, 0, 0), (3, 2, 0, 1, 0, 0, 0), (4, 2, 0, 0, 1, 1, 0), (5, 2, 0, 0, 1, 0, 1), (6, 2, 0, 0, 0, 1, 1) and (7, 3, 0, 0, 1, 1, 1) which are used in our numerical calculations, 
\begin{eqnarray}
{\mathcal Y}_1 &=& \sqrt{105 \over 32}{1\over \pi^2},  \nonumber\\
{\mathcal Y}_2 &=& \sqrt{385 \over 6}{3\over 16\pi^2}(3\cos(2\alpha_3)-1),\nonumber\\
{\mathcal Y}_3 &=& \sqrt{385 \over 2}{3\over 8\pi^2}\cos (2\alpha_2)\cos^2(\alpha_3), \nonumber\\
{\mathcal Y}_4 &=& -\sqrt{385\over 2}{3\over 4\pi^2}\cos\alpha_2\sin\alpha_2\cos^2\alpha_3\nonumber\\
               &\times& \left[\cos\theta_1 \cos\theta_2+\cos(\phi_1-\phi_2)\sin\theta_1\sin\theta_2\right], \nonumber\\
{\mathcal Y}_5 &=& -\sqrt{385\over 2}{3\over 4\pi^2}\cos\alpha_2\cos\alpha_3\sin\alpha_3\nonumber\\
               &\times& \left[\cos\theta_1 \cos\theta_3+\cos(\phi_1-\phi_3)\sin\theta_1\sin\theta_3\right], \nonumber\\
{\mathcal Y}_6 &=& -\sqrt{385\over 2}{3\over 4\pi^2}\sin\alpha_2\cos\alpha_3\sin\alpha_3\nonumber\\
               &\times& \left[\cos\theta_2 \cos\theta_3+\cos(\phi_2-\phi_3)\sin\theta_2\sin\theta_3\right], \nonumber\\
{\mathcal Y}_7 &=& i \sqrt{5005}{3\over 8\pi^2}\sin\alpha_2\cos\alpha_2\sin\alpha_3\cos^2\alpha_3\nonumber\\
               &\times& [\cos\theta_3\sin\theta_1\sin\theta_2\sin(\phi_1-\phi_2)\nonumber\\
               &-&\sin\theta_3\cos\theta_2\sin\theta_1\sin(\phi_1-\phi_3)\nonumber\\
               &+&\sin\theta_3\cos\theta_1\sin\theta_2\sin(\phi_2-\phi_3)]. 
\end{eqnarray}

\section{Computing the Potential Matrix}
\label{appendix2}

The difficulty to calculate the potential matrix element $V^{\kappa\kappa'}$ \eqref{vmatrix} is the integration over the eight angels $\Omega=(\alpha_2, \alpha_3, \theta_1, \phi_1, \theta_2, \phi_2, \theta_3, \phi_3)$. Let us first consider the potential between the two quarks, 
\begin{eqnarray}
V^{\kappa\kappa'}_{12} &=& \int V_{12}(|{\bf r}_2-{\bf r}_1|){\mathcal Y}_\kappa^*(\Omega){\mathcal Y}_{\kappa'}(\Omega)d\Omega\nonumber\\
                       &=& \int V_{12}(\sqrt{2\mu/m}\rho \sin \alpha_3) {\mathcal Y}_\kappa^*(\Omega){\mathcal Y}_{\kappa'}(\Omega) d\Omega\nonumber\\
                       &=& N \int (\sin \alpha_3)^{l_3+l'_3} (\cos \alpha_3)^{K_{2}+K'_{2}}\nonumber\\
                       &\times& P_{n_3}^{l_3+1/2, K_{2}+2}(\cos 2 \alpha_3) P_{n'_3}^{l'_3+1/2, K'_{2}+2}(\cos 2 \alpha_3)\nonumber\\
                       &\times& V_{12}(\sqrt{2\mu/m}\, \rho \sin \alpha_3) \mathrm{d}\alpha_3,
\end{eqnarray}
where $N$ is a trivial seven dimensional integration, and the integration over $\alpha_3$ can be done easily. The above reduction from eight to one dimensional integration comes from the fact that $|{\bf r}_2-{\bf r}_1|$ is only a function of $\alpha_3$, by the definition \eqref{jacobi}. For the other interaction between a quark and an antiquark or two antiquarks, there is no such a reduction, because in general case $|{\bf r}_j-{\bf r}_i|$ depends on more angels. One way to effectively reduce the dimensions of the integration is to make a rotation in the coordinate space to guarantee $|{\bf r}_j-{\bf r}_i| \sim \sin\tilde\alpha_3$. This rotation in coordinate space is equivalent to a particle index permutation. We extend the special Jacobi transformation \eqref{jacobi} to a general one, 
\begin{eqnarray}
{\bf x}_1^{(ij)}&=&\sqrt{3m\over 4\mu}\left({\bf r}_l-{{\bf r}_i+{\bf r}_j+{\bf r}_k \over 3}\right),\nonumber\\
{\bf x}_2^{(ij)}&=&\sqrt{2m \over 3\mu}\left({\bf r}_k-{{\bf r}_i+{\bf r}_j\over 2}\right),\nonumber\\
{\bf x}_3^{(ij)}&=&\sqrt{{m\over 2\mu}}\left({\bf r}_j-{\bf r}_i\right). 
\label{jacobi2}
\end{eqnarray}
The two groups of relative coordinates \eqref{jacobi} and \eqref{jacobi2} are connected via a transformation,
\begin{equation}
\left(\begin{array}{c}
{\bf x}_1^{(12)} \\
{\bf x}_2^{(12)} \\
{\bf x}_3^{(12)} \\
\end{array}\right)
= A^{(ij)}
\left(\begin{array}{c}
{\bf x}_1^{(ij)} \\
{\bf x}_2^{(ij)} \\
{\bf x}_3^{(ij)} \\
\end{array}\right),
\end{equation}
and the hyperspherical harmonic functions ${\mathcal Y}_\kappa(\Omega^{ij})$ corresponding to \eqref{jacobi2} are related to ${\mathcal Y}_\kappa(\Omega)$ to \eqref{jacobi} via a transformation,
\begin{equation}
{\mathcal Y}_{\kappa}(\Omega) = \sum_{\kappa'} R^{(ij)}_{\kappa \kappa'} {\mathcal Y}_{\kappa'}(\Omega^{ij}),
\end{equation}
where $R^{(ij)}_{\kappa \kappa'}$ are called Raynal-Revai coefficients~\cite{raynal1970transformation,jibuti1988construction}. 

With the Raynal-Revai matrix, any potential element $V_{ij}^{\kappa\kappa'}$ is simplified to the calculation of $V_{12}^{\kappa\kappa'}$, 
\begin{eqnarray}
V^{\kappa\kappa'}_{ij} &=& \int V_{ij} (|{\bf r}_j-{\bf r}_i|){\mathcal Y}_\kappa^*(\Omega) {\mathcal Y}_{\kappa'}(\Omega)d\Omega\nonumber\\
                       &=& \sum_{\omega\omega'} (R^{(ij)}_{\kappa\omega})^* R^{(ij)}_{\kappa'\omega'}\nonumber\\
                       &\times& \int V_{ij} (|{\bf r}_j-{\bf r}_i|){\mathcal Y}_\omega^*(\Omega^{ij}){\mathcal Y}_{\omega'}(\Omega^{ij})d\Omega^{ij}\nonumber\\
                       &=& \sum_{\omega\omega'} (R^{(ij)}_{\kappa\omega})^* R^{(ij)}_{\kappa'\omega'} V^{\omega\omega'}_{12}.
\end{eqnarray}

\end{appendix}



\begin{thebibliography}{20}

\bibitem{Mathieu:2008me}
V.~Mathieu, N.~Kochelev and V.~Vento,
Int. J. Mod. Phys. E \textbf{18}, 1(2009).

\bibitem{Ochs:2013gi}
W.~Ochs,
J. Phys. G \textbf{40}, 043001(2013).

\bibitem{Meyer:2015eta}
C.~A.~Meyer and E.~S.~Swanson,
Prog. Part. Nucl. Phys. \textbf{82}, 21(2015).

\bibitem{Chanowitz:1982qj}
M.~S.~Chanowitz and S.~R.~Sharpe,
Nucl. Phys. B \textbf{222}, 211(1983).

\bibitem{Esposito:2016noz}
A.~Esposito, A.~Pilloni and A.~D.~Polosa,
Phys. Rep. \textbf{668},1(2017).

\bibitem{Karliner:2017qhf}
M.~Karliner, J.~L.~Rosner and T.~Skwarnicki,
Annu. Rev. Nucl. Part. Sci. \textbf{68}, 17(2018).

\bibitem{DeRujula:1976zlg}
A.~De Rujula, H.~Georgi and S.~L.~Glashow,
Phys. Rev. Lett. \textbf{38}, 317(1977).

\bibitem{Guo:2017jvc}
F.~K.~Guo, C.~Hanhart, U.~G.~Meißner, Q.~Wang, Q.~Zhao and B.~S.~Zou,
Rev. Mod. Phys. \textbf{90}, 015004(2018).

\bibitem{Choi:2003ue}
S.~K.~Choi \textit{et al.} (Belle Collaboration),
Phys. Rev. Lett. \textbf{91}, 262001(2003).

\bibitem{Richard:2016eis}
J.~M.~Richard,
Few Body Syst. \textbf{57}, 1185(2016).

\bibitem{Hosaka:2016pey}
A.~Hosaka, T.~Iijima, K.~Miyabayashi, Y.~Sakai and S.~Yasui,
Prog. Theor. Exp. Phys. \textbf{2016} (2016)062C01.

\bibitem{Ali:2017jda}
A.~Ali, J.~S.~Lange and S.~Stone,
Prog. Part. Nucl. Phys. \textbf{97}, 123(2017).

\bibitem{Liu:2019zoy}
Y.~R.~Liu, H.~X.~Chen, W.~Chen, X.~Liu and S.~L.~Zhu,
Prog. Part. Nucl. Phys. \textbf{107}, 237(2019).

\bibitem{Badalian:1985es}
A.~M.~Badalian, B.~L.~Ioffe and A.~V.~Smilga,
Nucl. Phys. B \textbf{281}, 85(1987).

\bibitem{Ader:1981db}
J.~P.~Ader, J.~M.~Richard and P.~Taxil,
Phys. Rev. D \textbf{25}, 2370(1982).

\bibitem{Zouzou:1986qh}
S.~Zouzou, B.~Silvestre-Brac, C.~Gignoux and J.~M.~Richard,
Z. Phys. C \textbf{30}, 457(1986).

\bibitem{Brink:1998as}
D.~M.~Brink and F.~Stancu,
Phys. Rev. D \textbf{57}, 6778(1998).

\bibitem{Berezhnoy:2011xn}
A.~V.~Berezhnoy, A.~V.~Luchinsky and A.~A.~Novoselov,
Phys. Rev. D \textbf{86},  034004(2012).

\bibitem{Karliner:2016zzc}
M.~Karliner, S.~Nussinov and J.~L.~Rosner,
Phys. Rev. D \textbf{95}, 034011(2017).

\bibitem{Debastiani:2017msn}
V.~R.~Debastiani and F.~S.~Navarra,
Chin. Phys. C \textbf{43}, 013105(2019).

\bibitem{Wang:2019rdo}
G.~J.~Wang, L.~Meng and S.~L.~Zhu,
Phys. Rev. D \textbf{100}, 096013(2019).

\bibitem{Liu:2019zuc}
M.~S.~Liu, Q.~F.~Lü, X.~H.~Zhong and Q.~Zhao,
Phys. Rev. D \textbf{100}, 016006(2019).

\bibitem{Chen:2020lgj}
X.~Chen,
[arXiv:2001.06755].

\bibitem{Yang:2020rih}
G.~Yang, J.~Ping, L.~He and Q.~Wang,
[arXiv:2006.13756].

\bibitem{Lu:2020cns}
Q.~F.~Lü, D.~Y.~Chen and Y.~B.~Dong,
Eur. Phys. J. C \textbf{80}, 871(2020).

\bibitem{Bicudo:2015vta}
P.~Bicudo, K.~Cichy, A.~Peters, B.~Wagenbach and M.~Wagner,
Phys. Rev. D \textbf{92}, 014507(2015).

\bibitem{Bicudo:2017usw}
P.~Bicudo, M.~Cardoso, O.~Oliveira and P.~J.~Silva,
Phys. Rev. D \textbf{96}, 074508 (2017).

\bibitem{Chen:2016jxd}
W.~Chen, H.~X.~Chen, X.~Liu, T.~G.~Steele and S.~L.~Zhu,
Phys. Lett. B \textbf{773}, 247(2017).

\bibitem{Wang:2017jtz}
Z.~G.~Wang,
Eur. Phys. J. C \textbf{77}, 432(2017).

\bibitem{Aaij:2020fnh}
R.~Aaij \textit{et al.} (LHCb Collaboration),
[arXiv:2006.16957].

\bibitem{Adamczyk:2014uip}
L.~Adamczyk \textit{et al.} (STAR Collaboration),
Phys. Rev. Lett. \textbf{113}, 142301 (2014).

\bibitem{Abelev:2012vra}
B.~Abelev \textit{et al.} (ALICE Collaboration),
J. High Energy Phys. \textbf{07} (2012) 191. 

\bibitem{Bedjidian:2004gd}
M.~Bedjidian, et al.,
[arXiv:hep-ph/0311048].

\bibitem{Zhao:2020jqu}
J.~Zhao, K.~Zhou, S.~Chen and P.~Zhuang,
Prog. Part. Nucl. Phys. \textbf{114}, 103801 (2020).

\bibitem{Fries:2003kq}
R.~J.~Fries, B.~Muller, C.~Nonaka and S.~A.~Bass,
Phys. Rev. C \textbf{68}, 044902 (2003).

\bibitem{Molnar:2003ff} 
  D.~Molnar and S.~A.~Voloshin,
  Phys.\ Rev.\ Lett.\  {\bf 91}, 092301 (2003).
  
\bibitem{Lin:2002rw} 
  Z.~W.~Lin and C.~M.~Ko,
  Phys.\ Rev.\ Lett.\  {\bf 89}, 202302 (2002).  

\bibitem{Hwa:2002tu} 
  R.~C.~Hwa and C.~B.~Yang,
  Phys.\ Rev.\ C {\bf 67}, 034902 (2003).  
  
\bibitem{Oh:2009zj} 
  Y.~Oh, C.~M.~Ko, S.~H.~Lee, and S.~Yasui,
  Phys.\ Rev.\ C {\bf 79}, 044905 (2009).  

\bibitem{Caswell:1985ui} 
  W.~E.~Caswell and G.~P.~Lepage,
  Phys.\ Lett.\  {\bf 167B}, 437 (1986).

\bibitem{Brambilla:1999xf} 
  N.~Brambilla, A.~Pineda, J.~Soto, and A.~Vairo,
  Nucl.\ Phys.\ B {\bf 566}, 275 (2000).  

\bibitem{Satz:2005hx} 
  H.~Satz,
  J.\ Phys.\ G {\bf 32}, R25 (2006). 
  
\bibitem{Zhao:2017gpq}
J.~Zhao and P.~Zhuang,
Few Body Syst. \textbf{58}, 100(2017).

\bibitem{Shi:2019tji}
S.~Shi, J.~Zhao and P.~Zhuang,
Chin. Phys. C \textbf{44}, 8 (2020).

\bibitem{Zhao:2016ccp}
J.~Zhao, H.~He and P.~Zhuang,
Phys. Lett. B \textbf{771}, 349(2017).

\bibitem{He:2014tga}
H.~He, Y.~Liu and P.~Zhuang,
Phys. Lett. B \textbf{746}, 59(2015).

\bibitem{Wong:2001td}
C.~Y.~Wong, E.~S.~Swanson and T.~Barnes,
Phys. Rev. C \textbf{65}, 014903(2001).

\bibitem{Kawanai:2011jt} 
  T.~Kawanai and S.~Sasaki,
  Phys.\ Rev.\ D {\bf 85}, 091503 (2012).   
  
\bibitem{Krivec:1998}
R. Krivec,
Few-Body Syst. 25, 199 (1998).  

\bibitem{Park:2013fda}
W.~Park and S.~H.~Lee,
Nucl. Phys. A \textbf{925}, 161(2014).

\bibitem{Barnea:1999be}
N.~Barnea, W.~Leidemann and G.~Orlandini,
Phys. Rev. C \textbf{61}, 054001(2000).

\bibitem{Barnea:2006sd}
N.~Barnea, J.~Vijande and A.~Valcarce,
Phys. Rev. D \textbf{73}, 054004 (2006).

\bibitem{H.W. Crater} 
 H.W. Crater, J. Comput. Phys. 115,  470(1994).
 
\bibitem{SSZ} 
S.~Shi, J.~Zhao, and P.~Zhuang. (to be published).

\bibitem{Varga:1995dm} 
K.Varga and Y.Suzuki, Phys. Rev. {\bf C52}, 2885(1995).

\bibitem{Suzuki:1998bn}
Y.Suzuki and K.Varga, Lect. Notes Phys. Monogr. {\bf 54}, 1(1998). 

\bibitem{SilvestreBrac:2007sg} 
B.Silvestre-Brac and V.Mathieu, Phys. Rev. {\bf E76}, 046702(2007). 

\bibitem{SilvestreBrac:2008} 
B.Silvestre-Brac and V.Mathieu, Phys. Rev. {\bf E77}, 036706(2008).  
  
\bibitem{Hiyama:2003cu}
E.~Hiyama, Y.~Kino and M.~Kamimura,
Prog. Part. Nucl. Phys. \textbf{51}, 223(2003).

\bibitem{Zyla:2020zbs}
P.~A.~Zyla \textit{et al.} [Particle Data Group],
Prog. Theor. Exp. Phys. \textbf{2020}, 083C01 (2020).

\bibitem{Bazavov:2011nk} 
  A.~Bazavov {\it et al.},
  Phys.\ Rev.\ D {\bf 85}, 054503 (2012).
  
\bibitem{Fodor:2004nz}
Z.~Fodor and S.~D.~Katz,
J. High Energy Phys. \textbf{04} (2004), 050.

\bibitem{Xin:2014ela}
X.~Y.~Xin, S.~X.~Qin and Y.~X.~Liu,
Phys. Rev. D \textbf{90}, 076006(2014).

\bibitem{Li:2018ygx}
Z.~Li, K.~Xu, X.~Wang and M.~Huang,
Eur. Phys. J. C \textbf{79}, 245 (2019).

\bibitem{Petreczky:2010yn} 
  P.~Petreczky,
  J.\ Phys.\ G {\bf 37}, 094009 (2010).  
  
\bibitem{Lafferty:2019jpr}
D.~Lafferty and A.~Rothkopf,
Phys. Rev. D \textbf{101}, 056010 (2020).

\bibitem{Guo:2012hx}
X.~Guo, S.~Shi and P.~Zhuang,
Phys. Lett. B \textbf{718}, 143(2012).

\bibitem{Greco:2003vf} 
  V.~Greco, C.~M.~Ko and R.~Rapp,
  Phys.\ Lett.\ B {\bf 595}, 202 (2004).     

\bibitem{Cho:2017dcy}
S.~Cho \textit{et al.} (ExHIC Collaboration),
Prog. Part. Nucl. Phys. \textbf{95}, 279(2017).

\bibitem{Fontoura:2019opw}
C.~E.~Fontoura, G.~Krein, A.~Valcarce and J.~Vijande,
Phys. Rev. D \textbf{99}, 094037 (2019).

\bibitem{Abelev:2012sea}
B.~Abelev \textit{et al.} (ALICE Collaboration),
Eur. Phys. J. C \textbf{73}, 2456 (2013).

\bibitem{Cacciari:2012ny}
  M.~Cacciari, S.~Frixione, N.~Houdeau, M.~L.~Mangano, P.~Nason and G.~Ridolfi,
  J. High Energy Phys. {\bf 10} (2012) 137.

\bibitem{Kolb:2003dz}
P.~F.~Kolb and U.~W.~Heinz,
[arXiv:nucl-th/0305084].
  
\bibitem{Sollfrank:1996hd}
J.~Sollfrank, P.~Huovinen, M.~Kataja, P.~V.~Ruuskanen, M.~Prakash and R.~Venugopalan,
Phys. Rev. C \textbf{55}, 392(1997).

\bibitem{Abelev:2013lca}
B.~Abelev \textit{et al.} (ALICE Collaboration),
Phys. Rev. Lett. \textbf{111}, 102301(2013).

\bibitem{Aaij:2011yc}
R.~Aaij \textit{et al.} (LHCb Collaboration),
Phys. Lett. B \textbf{707}, 52(2012).

\bibitem{Khachatryan:2016ydm}
V.~Khachatryan \textit{et al.} (CMS Collaboration),
J. High Energy Phys. \textbf{05} (2017), 013.

\bibitem{jibuti1988construction}
  R.~I.~Jibuti, N.~B.~Krupennikova, and L.~L.~Sarkisyan,
  Few-Body Syst. \textbf{4}, 151 (1988).

\bibitem{raynal1970transformation}
  J.~Raynal, and J.~Revai,
 Il Nuovo Cimento A (1965-1970) \textbf{68}, 612(1970).

\bibitem{Becchi:2020mjz}
C.~Becchi, A.~Giachino, L.~Maiani and E.~Santopinto,
Phys. Lett. B \textbf{806}, 135495(2020).

\bibitem{Becchi:2020uvq}
C.~Becchi, A.~Giachino, L.~Maiani and E.~Santopinto,
[arXiv:2006.14388].

\end{thebibliography}
\end{document}